\documentclass[]{emulateapj}

\newcommand\coplus{CO$^+$~}

\newcommand\ohplus{OH$^+$~}

\newcommand\hcoplus{HCO$^+$~}

\newcommand\hhs{H$_2$S~}
\newcommand\hhhplus{H$_3^+$~}

\newcommand\hho{H$_2$O~}

\newcommand\hh{H$_2$~}

\newcommand\soo{SO$_2$~}

\newcommand\hhco{H$_2$CO~}

\newcommand{\tto}[2]{$#1 \times 10^{#2}$}

\shorttitle{Chemical Modelling of Young Stellar Objects, I. Method and Benchmarks}
\shortauthors{S. Bruderer et al.}

\begin{document}

\title{Chemical Modelling of Young Stellar Objects, I. Method and Benchmarks}

\author{S. Bruderer\altaffilmark{1}}
\altaffiltext{1}{Institute of Astronomy, ETH Zurich, CH-8093 Zurich, Switzerland}
\email{simonbr@astro.phys.ethz.ch}

\author{S. D. Doty\altaffilmark{2}}
\altaffiltext{2}{Department of Physics and Astronomy, Denison University, Granville, OH 43023, USA}

\author{A.O. Benz\altaffilmark{1}}

\begin{abstract}
Upcoming facilities such as the Herschel Space Observatory or ALMA will deliver a wealth of molecular line observations of young stellar objects (YSOs). Based on line fluxes, chemical abundances can then be estimated by radiative transfer calculations. To derive physical properties from abundances, the chemical network needs to be modeled and fitted to the observations. This modeling process is however computationally exceedingly demanding, particularly if in addition to density and temperature, far UV (FUV) irradiation, X-rays, and multi-dimensional geometry have to be considered.\\
We develop a fast tool, suitable for various applications of chemical modeling in YSOs. A grid of the chemical composition of the gas having a density, temperature, FUV irradiation and X-ray flux is pre-calculated as a function of time. A specific interpolation approach is developed to reduce the database to a feasible size. Published models of AFGL 2591 are used to verify the accuracy of the method. A second benchmark test is carried out for FUV sensitive molecules.\\
The novel method for chemical modeling is more than 250,000 times faster than direct modeling and agrees within a mean factor of 1.35. The tool is distributed for public use. Main applications are ({\it i}) fitting physical parameters to observed molecular line fluxes and ({\it ii}) derive chemical abundances for 2D and 3D models. They will be presented in two future publications of this series.\\
In the course of devloping the method, the chemical evolution is explored: We find that X-ray chemistry in envelopes of YSOs can be reproduced by means of an enhanced cosmic-ray ionization rate with deviations less than $25\%$, having the observational consequence that molecular tracers for X-rays are hard to distinguish from cosmic ray ionization tracers. We provide the detailed prescription to implement this total ionization rate approach in any chemical model. We further find that the abundance of CH$^+$ in low-density gas with high ionization can be enhanced by the recombination of doubly ionized carbon (C$^{++}$) and suggest a new value for the initial abundance of the main sulphur carrier in the hot-core.
\end{abstract}


%
%
\section{Introduction} \label{sec:intro}

There is an interesting but little explored phase in star formation when the cloud core collapses, but the protostar is still deeply embedded. In this phase, the temperature of the interior envelope exceeds 100 K, outflows are observed, protoplanetary disks form, and the protostar begins to radiate in UV and X-rays. These physical processes can best be observed in low temperature lines, some atomic but most molecular. A wealth of new line data of young stellar objects (YSOs) will be available soon by upcoming facilities like the Herschel Space Observatory or the Atacama Large Millimeter Array (ALMA, e.g. \citealt{vanDishoeck08a} for a recent review). Two steps are necessary to constrain physical parameters by molecular line observations: \textit{(i)} Radiative transfer calculations are applied to estimate the abundance of the observed molecule, and \textit{(ii)} modeling the chemical network relates the abundance to physical parameters such as age, density, temperature, far UV (FUV) and X-ray irradiation by the protostar. Although the present knowledge of these networks is still limited by unknown reactions on grain surfaces, in some cases it is possible to derive chemical abundances from a set of initial physical parameters. The abundances form the bases for radiative transfer calculations. The resulting line fluxes can then be compared to the observed fluxes. Finally, the input physical parameters for the chemical modeling are changed until the derived line fluxes fit the observed ones, for instance producing a minimum in a $\chi^2$ test. Chemical modeling simulates a network of molecular species reacting with each other (e.g. \citealt{Doty02}). Contrary to an atomic gas, molecular abundances are not conserved, but depend on local physical parameters. In chemical simulations, the abundances of different species are related to the change of one species in a set of non-linear coupled differential equations. Solving these equations is a time-consuming task for a chemical network consisting of several hundred species connected by thousands of reactions.\\

While early astrochemical models of envelopes of YSOs assumed a fixed density and temperature (e.g. \citealt{Leung84}), more recent models considered space-dependent chemistry (e.g. \citealt{Caselli93} or \citealt{Millar97a}). So far, chemical models have only been fitted to observations under the assumption of spherically symmetric, one-dimensional physical parameters (\citealt{Doty02}; \citealt{Staeuber05}). This is a questionable assumption as YSOs have inherently 2D or 3D geometries with outflow cavities, disks etc. For such geometries the number of cells for which chemical modeling has to be performed increases from about one hundred to $10^4$ and more. Fitting requires repetition of many similar calculations. It becomes inefficient and beyond present resources for 2D or 3D geometries and for a large sample of sources.\\

In a dynamical situation, the ``chemical age'' (evolution time since the start from an initial composition) may be different from the age of the YSO. The chemistry in a gravitationally collapsing envelope has been studied by \citet{Rodgers03} and \citet{Lee04}. \citet{Doty06} have constructed a model including infall as well as the evolution of the central stellar object. They conclude that the chemical evolution of an ice-evaporated hot core appears to be younger than the time since the formation of the YSO by a factor of 4 to 10. ``Pseudo time-dependent'' models keep the physical structure constant with time, while the chemical abundances evolve. Such models nevertheless are reasonable approximations if the chemical composition evolves rapidly compared to the change of the relevant physical parameters, like for example in regions with strong FUV irradiation.\\

Undoubtedly, the ionization of molecular hydrogen, H$_2$, followed by fast ion-molecule reactions involves extremely rapid processes. Ionized molecules are a driving force of the chemistry in the envelope of YSOs (\citealt{Dalgarno06}). Cosmic rays account for ionizations mainly in the outer part of the envelope. In the inner part, high-energy radiation from the YSO may be the dominant source of ionization. Accretion phenomena and shocks may account for FUV radiation in low-mass stars, while hot high-mass stars emit copious FUV photons mostly from their hot photosphere. Direct photoionization through FUV photons leads to a particular chemistry as observed in photo dissociation regions (PDRs). FUV radiation is attenuated by dust (e.g. \citealt{Montmerle01}). Unless a low-density region allows FUV radiation to escape from the innermost region of the envelope, its influence on the chemistry is restricted to a small volume behind the irradiated surface (\citealt{Staeuber04}). The observations of \citet{Staeuber07} revealed a much larger amount of several FUV sensitive molecules than predicted by their spherically symmetric models. These authors suggested outflows acting as paths for FUV photons to escape and penetrate the high-density gas in the border region between the outflow and the envelope. A multidimensional geometry is needed to model these ``outflow-walls''.\\

In addition to FUV, young stars are known to be strong emitters of X-rays with luminosities up to 10$^{32}$~ erg s$^{-1}$ in the 1 - 100 keV band (e.g. \citealt{Preibisch05a}). Magnetic activity is believed to be the origin of the X-rays, but the exact mechanism remains unclear. X-ray photons have a smaller absorption cross-section than FUV ($\propto \lambda^3$) and penetrate deeper into dense material than FUV. X-ray ionization thus decreases mostly by geometric dilution, $\propto r^{-2}$ with distance $r$ to the source. Various studies on the influence of X-ray irradiation on the chemistry of molecular clouds have been carried out (\citealt{Krolik83}, \citealt{Maloney96}, \citealt{Lepp96}, \citealt{Yan97}, \citealt{Staeuber05}). For low-mass YSOs, the presence of X-rays has implications on the disc and thus planet formation (e.g. \citealt{Ilgner08,Meijerink08}). In an early stage of star formation, FUV and X-ray radiation are absorbed by the large opacity of the envelope and are therefore not directly observable. The point of evolution when FUV and X-ray activity sets in is unknown.\\

The goal of our work is to provide a fast and simple method for chemical modeling. A large set of pre-calculated abundances for different ages and physical conditions (density, temperature, etc.) is used to quickly interpolate the chemical abundances for individual conditions in a model with physical conditions that change with position. The main problem is to reduce the high-dimensional space of physical parameter to a size that fits current hardware resources, but keep the accuracy of the interpolation at an acceptable level. Based on this new method for chemical modeling, observed data can be quickly fitted to physical parameters such as the chemical age or the X-ray luminosity. Also detailed 2D or 3D models may be constructed quickly allowing to study the influence of the geometry on observable parameters (Bruderer et al., in prep.)\\

This paper is organized as follows: In the first section, we describe the grid approach. We discuss the relevant parameters for the chemical composition and the similarity between X-ray and cosmic ray induced chemistry. In Section \ref{sec:chemmod}, the chemical model (network and reaction rates) used for this work is briefly described. Two benchmark tests on realistic applications are carried out and discussed in the following section. Section \ref{sec:application} describes the application of the method. In the following parts of this series of papers, we will present a detailed multidimensional chemical model of a high-mass star forming region applied to the enigmatic \coplus molecule and a parameter study of 2D effects on the interpretation of line observations as expected to be observed with Herschel and ALMA. \\

%
%
\section{A grid of chemical models} \label{sec:grid}

Chemical models solve for the molecular abundances using the rate equations 
\begin{equation} \label{eq:rate}
\frac{dn(i)}{dt}=\sum_{j} \, k_{i,j} \cdot n(j) + \sum_{j,k} \, k'_{i,j,k} \cdot n(j) \cdot n(k) + S(i)
\end{equation}
which relate the temporal evolution of a species, labeled by $i$, to the number density $n(j)$, $n(k)$~ [cm$^{-3}$] of some species $j$ and $k$. The constants of proportionality $k_{i,j}$ [s$^{-1}$] and $k'_{i,j,k}$~ [cm$^{3}$ s$^{-1}$] depend on physical properties like temperature, cosmic-ray ionization rate or FUV flux. Some applications require an additional source term $S(i)$ to account e.g. for molecules evaporating from dust grains or spatial flows.\\

Many of the reactions relevant in the interstellar environment at low temperature are difficult or impossible to measure in terrestrial conditions. In standard networks for astrochemical applications, a majority of reaction rates are thus only known to within a factor of two (\citealt{Woodall07}). Furthermore, uncertainties in the rate coefficients may grow due to the non-linear nature of the rate equations (Eq. \ref{eq:rate}). Studies on the quantitative uncertainty of the chemical abundances have been carried out only recently by \citet{Wakelam05a,Wakelam06b} and \citet{Vasyunin08}. Considering their results, it seems reasonable to require the interpolated abundances to agree within a factor of two with the fully calculated values.\\

The basis of our method is to interpolate abundances from a grid of physical parameters relevant for the chemical composition. These physical parameters form a multidimensional space, where the dimensions corresponds to temperature, density, cosmic-ray ionization rate and several parameters for the impact of high-energetic radiation (X-rays and FUV irradiation). For each grid point and all molecules in the chemical network, the temporal evolution is stored in a database. Therefore, time enters as an additional dimension. In order to explore a wide range in parameter space, we assume a rectangular grid: For every grid point along one axis, all possible combinations of grid points along all other axes are calculated. The total number of models is obtained by multiplying the number of grid points along each dimension. Thus, it is crucial to keep the number of dimensions and grid points as low as possible, while it must still be sufficient to guarantee the required interpolation accuracy.\\

\begin{figure*}[tbh]
\begin{center}
\includegraphics[width=0.7\textwidth]{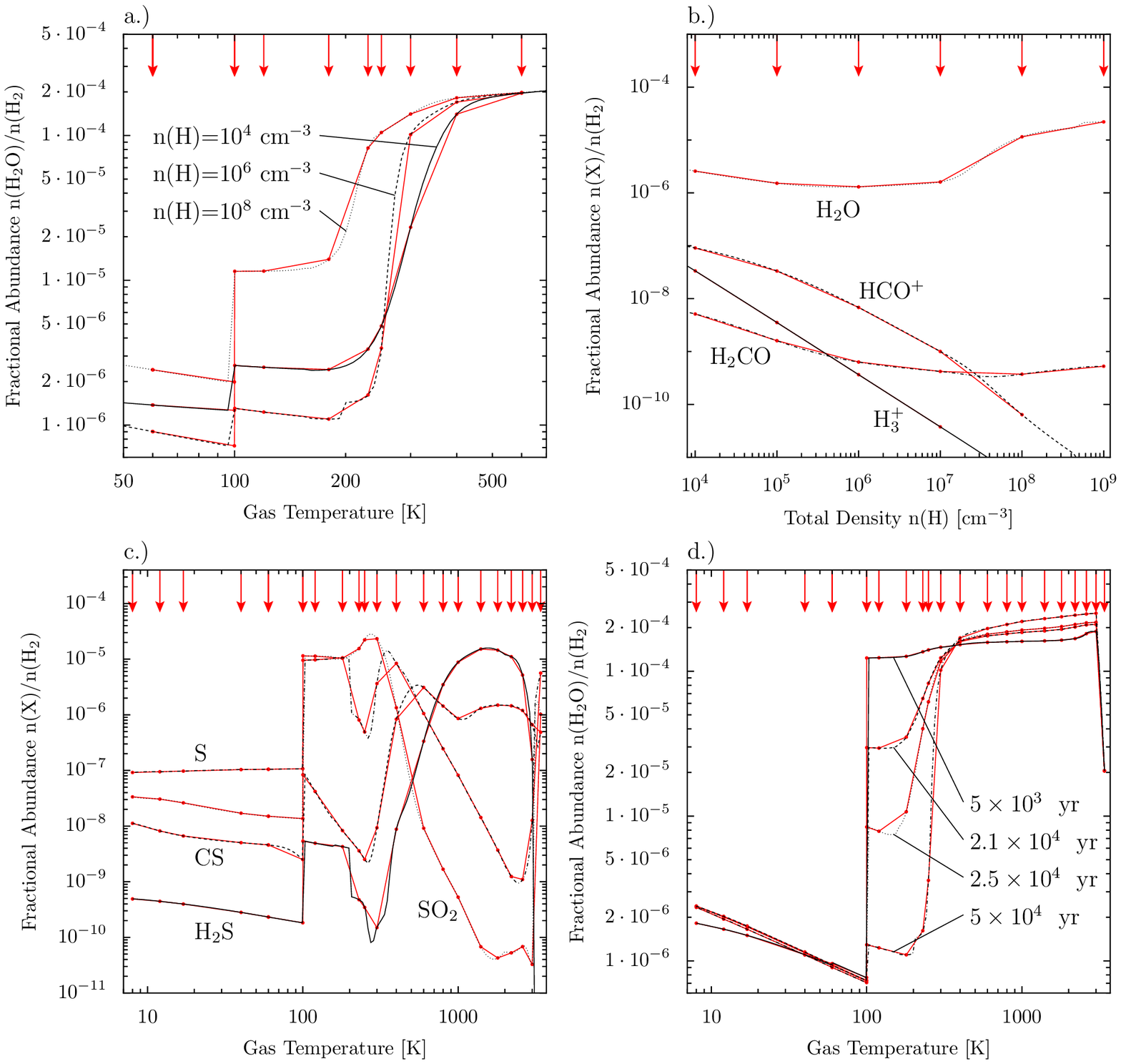}
\caption{Fractional abundance of different species depending on physical properties (temperature, density). The arrows indicate the positions of the interpolation points. The red lines show the results of the interpolation in log-log space. \textbf{a.)} Water in a region with X-ray irradiation versus temperature for three different densities. \textbf{b.)} Density dependence of four selected species related to the ionization fraction. \textbf{c.)} Sulphur bearing species in the temperature range of the grid. \textbf{d.)} X-ray irradiated water at different chemical ages versus temperature.\label{fig:slides_plot}}
\end{center}
\end{figure*}

As an example for problems to be met by the interpolation, we show the water fractional abundance with respect to \hh in Fig. \ref{fig:slides_plot}a: A parcel of gas and dust at a gas density between 10$^4$~ cm$^{-3}$~ and 10$^8$~ cm$^{-3}$~ irradiated by X-rays has been modeled in a similar way as \citet{Staeuber07}. For this figure, the fractional abundances are given at the time (the so called chemical age) of \tto{5}{4} yr according to the age of the high-mass protostellar object AFGL 2591 (\citealt{Staeuber05}).\\

The presentation of Fig. \ref{fig:slides_plot} with logarithmic scales in both dimensions suggests to use a linear interpolation of the fractional abundance in log-log space. The straight (red/grey) lines in the figure show interpolated fractional abundances. The temperatures marked by arrows at the top of the figure act as interpolation points. Two temperatures close to 100 K have been used, since water is assumed to evaporate at this temperature from dust grains. For an accurate interpolation in the temperature range between $\approx$200 and 400 K, a couple of interpolation points is required. The reaction OH + \hh $\rightarrow$~ \hho + H with a high activation energy becomes important in this temperature range, leading to little net variation in the water abundance at $T$  $>$ 400 K, while at lower temperature \hho is destroyed by X-rays \citep{Staeuber06}.\\

%
%

\subsection{Temperature and density dependence} \label{sec:tempdens}

The dependence of abundance on parameters can be much simpler than the temperature sensitivity of water: The fractional abundances of H$_3$, HCO$^+$, \hho and \hhco depending on the total gas density are shown in Fig. \ref{fig:slides_plot}b. For that plot, a temperature slightly above the water evaporation temperature of 100 K has been assumed. No X-ray or FUV irradiation is considered, but cosmic rays ionize \hh and account for the production of H$_3^+$. A cosmic-ray ionization rate $\zeta_{\rm c.r.}$ = \tto{4}{-16} s$^{-1}$, found by \citet{Doty04} in IRAS 16293-2422 is used for this plot. Higher values than the ``standard'' value of a few times 10$^{-17}$ s$^{-1}$ have also been suggested by \citet{vdTak00b}. This shows the necessity to include the cosmic-ray ionization rate as a dimension for the interpolation approach. The rate of ionizations by cosmic ray particles and thus the density of \hhhplus is independent of the H$_2$ density. Therefore, the fractional abundance of \hhhplus is approximately inversely proportional to the density. The same effect is found for HCO$^+$, produced by the reaction \hhhplus + CO $\rightarrow$ \hcoplus + H$_2$. In contrast, water and formaldehyde are not directly related to the ionization rate and their fractional abundances does not vary more than an order of magnitude in the explored density range.\\

As can be seen in Fig. \ref{fig:slides_plot}a and \ref{fig:slides_plot}b, the fractional abundance is a much stronger function of temperature than of density. This is due to the exponential dependence of some reaction rates on temperature. Reactions between neutrals that do not involve radicals or atoms often have considerable activation barriers. Their reactions rate is proportional to the Boltzmann factor $\exp(-E_a/k_b T)$ in the usual Arrhenius expression, where $E_a$ denotes the activation energy, $k_B$ is the Boltzmann constant and $T$ the kinetic temperature of the gas.\\

Selecting grid points in temperature is most important for the accuracy of the interpolation. In addition to water, sulphur bearing species also pose problems: Many reactions have a large activation energy and their fractional abundance is thus expected to be especially dependent on temperature. Indeed, Fig. \ref{fig:slides_plot}c shows a strong dependence of H$_2$S, CS, \soo and atomic sulphur on temperature. Again, we overplot interpolated abundances onto the fully calculated results. For temperatures below 100 K, only a small number of grid points is needed, but the hot-core regime of $T$ $>$ 100 K requires a much larger amount in order to keep deviations small.\\

How can we select the grid points along the temperature axis of the grid? Two methods have been tested: ({\it i}) an unbiased method and ({\it ii}) hand-placed points. For method ({\it i}), the grid points are chosen based on the number of reactions with activation temperature in a certain temperature range. The activation temperature of a reaction is obtained by scaling to the activation energy of the very important reaction OH + \hh $\rightarrow$ \hho + H which proceeds at a gas temperature above $\approx$ 250 K (Fig. \ref{fig:slides_plot}a and \citealt{Staeuber06}). The grid points are then placed to properly sample temperature ranges with higher activation temperatures. For method ({\it ii}), the grid points are placed by hand based on parameter studies like Fig. \ref{fig:slides_plot}a. The selection of species given by \citet{Staeuber05} and other important molecules are considered at different chemical ages.\\

The interpolation quality of the two methods has been tested using the technique presented in Section \ref{sec:comp_afgl}. We find a considerably better interpolation accuracy using the grid with hand-placed grid points for the following reason: The unbiased approach does not take into account the relative importance of different reactions: e.g. the reaction between OH and \hh, despite its substantial influence on the chemical network since the key molecule water is involved. As Fig. \ref{fig:slides_plot}a shows for this reaction, it is not sufficient to put two grid points near 250 K in order to obtain a good interpolation quality. The temporal evolution of the water abundance reveals another difficulty in the selection of the grid points (Fig. \ref{fig:slides_plot}d): For chemical ages between a few times $10^3$ and \tto{5}{4} yr, the fractional abundance cannot be interpolated well using the same set of grid points: The gradient of the water abundance for $100 <$ $T$ $< 250$ K and $T$ $>$ 250 K changes with the chemical evolution. Thus, the base point for the connection between the low and high water abundance shifts from 100 K to 250 K.\\

A similar effect is also observed in Fig. \ref{fig:slides_plot}a, where different total densities lead to different timescales of the water destruction. While the fractional abundance of water varies by approximately two orders of magnitude, species connected to its network may be affected even stronger. For example the fractional abundance of H$_2$S having its main destroyer (HCO$^+$) in common with \hho increases by many orders of magnitude between 250 K and 600 K (Fig. \ref{fig:slides_plot}c). We finally decided to implement 23 hand picked points at 8, 12, 17, 40, 59.9, 60.1, 99.9, 100.1, 120, 180, 230, 250, 300, 400, 600, 800, 1000, 1400, 1800, 2200, 2600, 3000 and 3400 K for our work. This range thus covers conditions from cold dark clouds to hot PDR like regions, heated by FUV photons.\\

The selection of the grid points for the density axis is less critical due to the smooth abundance profile along this dimension. We implement one grid point per order of magnitude in density. The density is taken to be within $10^4$ - $10^9$ cm$^{-3}$, sufficient to model envelopes of low-mass and high-mass YSOs (\citealt{Jorgensen02}, \citealt{Maret02}, \citealt{vdTak00a}).

%
%
\begin{table*}
\begin{center}
\caption{Parameters for the grid of chemical models, their physical range in the envelope of YSOs and the required number of models to achieve a good interpolation accuracy ($\approx$ a factor of two for most species/physical conditions). The parameters are defined and described in the text.\label{tab:param_grid}}
\begin{tabular}{l l c}
\tableline\tableline
Parameter & Range & Grid points \\
\tableline
\multicolumn{3}{l}{\textit{straight-forward implementation:}}\\
Density $n$                                                  & $10^4 - 10^9$ cm$^{-3}$        & 6 \\
Temperature  $T$                                             & $8 - 3400$ K                   & 23\\
X-ray flux  $F_{\rm X}$                                      & $10^{-6} - 10$ erg cm$^{-2}$ s$^{-1}$ & 14\\
Plasma temperature\tablenotemark{a}  $T_{\rm X}$             & $10^7 - 3 \times 10^8$ K       & 4\\
X-ray Attenuation\tablenotemark{a}  $N({\rm H}_{\rm tot})$   & $10^{21} - 10^{24}$ cm$^{-2}$  & 6\\
Cosmic-ray ion. rate  $\zeta_{cr}$                           & $10^{-17} - 10^{-15}$ s$^{-1}$ & 5 \\
FUV flux  $G_0$                                              & $0 - 10^7$ ISRF                & 9 \\
FUV attenuation\tablenotemark{a}  $\tau$                     & 0.1 - 10 A$_{\rm V}$           & 10 \\
                                                             &                                & $1.7\times10^7$ \\
\multicolumn{3}{l}{\textit{improved implementation:}}\\
Density  $n$                                                 & $10^4 - 10^9$ cm$^{-3}$        & 6 \\
Temperature  $T$                                             & $8 - 3400$ K                   & 23\\
Total ionization rate                                        &                                & \\
$\zeta_{\rm tot}$($\zeta_{cr}$,
       $F_{\rm X}$,$T_{\rm X}$,$N({\rm H}_{\rm tot})$)       & $10^{-17} - 10^{-12}$ s$^{-1}$ & 11 \\
$\alpha$($G_0$,$\tau$)                                       & arbitrary units                & 15 \\
$\beta$($G_0$,$\tau$)\tablenotemark{a}                       & arbitrary units                & 5 \\
                                                             &                                & $1.1\times10^5$ \\
\tableline
\end{tabular}
\tablenotetext{a}{The models for $G_0$=0 or $F_{\rm X}$=0 do not depend on the attenuating column density and plasma temperature. The same simplification is possible for high values of $\alpha$(G$_0$,$\tau$).}
\end{center}
\end{table*}

%
%
\subsection{FUV driven chemistry} \label{sec:uvdriven}

Reaction rates for photodissociation or ionization of molecules by FUV radiation can be written as (e.g. \citealt{vanDishoeck06c})
\begin{equation} \label{eq:uv_rate_integral}
k = \int_{E_{\rm min}}^{E_{\rm max}} J(E) \sigma(E) dE \ \ \ {\rm [s}^{-1}{\rm ]} \ .
\end{equation}
The mean intensity of the FUV radiation field $J(E)$ [cm$^{-2}$ s$^{-1}$ erg$^{-1}$] times the cross-section for photoabsorption and ionization $\sigma(E)$ [cm$^{2}$] is integrated between the average dust work function ($E_{\rm min} \approx 6$ eV) and the ionization energy of hydrogen ($E_{\rm max}=13.6$~ eV). For these photon energies, photodissociation proceeds through line absorption and absorption of the continuum by dust. Line absorption occurs by other species or by the considered species itself (so called self-shielding). In principle, the exact shape of the spectra is thus required to calculate the chemical rates. As an approximation, \citet{vanDishoeck88} used the spectral shape of the interstellar radiation field (ISRF, e.g. \citealt{Habing68} or \citealt{Draine78}) and fitted rate coefficients to the equation 
\begin{equation} \label{eq:uv_rate_fit}
k=G_0 \cdot C \cdot \exp\left(-\gamma \cdot \tau \right) \ \ \ {\rm [s}^{-1}{\rm ]} \ ,
\end{equation}
where $G_0$ is a scaling factor to the interstellar radiation field and $\tau = A_v/1.086$, $A_V$ being the extinction at visual wavelengths due to dust. It is calculated from the total hydrogen column density by the conversion factor $N({\rm H}_{\rm tot}) = 2 N({\rm H}_2) + N({\rm H}) \approx 1.87 \times 10^{21} \cdot A_v \ {\rm cm}^{-2}$ (\citealt{Bohlin78} for a dust reddening of $R_v \approx 3.1$). The values of the constants C and $\gamma$ in Eq. (\ref{eq:uv_rate_fit}) depend on the spectrum, used to for the fitting. For hot FUV sources with $T > 20\,000$ K, it is safe to use fits to the interstellar radiation field, dominated by O and B stars. $G_0$ is then scaled with respect to the average interstellar flux of \tto{1.6}{-3} erg cm$^{-2}$ s$^{-1}$ at photon energies between 6 and 13.6 eV. For colder sources such as accretion hot spots, the intensity of the spectra decreases at lower wavelengths, and C and $\gamma$ should be adjusted (see \citealt{vanDishoeck08b}). The value of $\gamma$ depends furthermore on the adopted dust model as well as on the range of $A_v$ used for the fitting.\\

\begin{figure}[ht]
\plotone{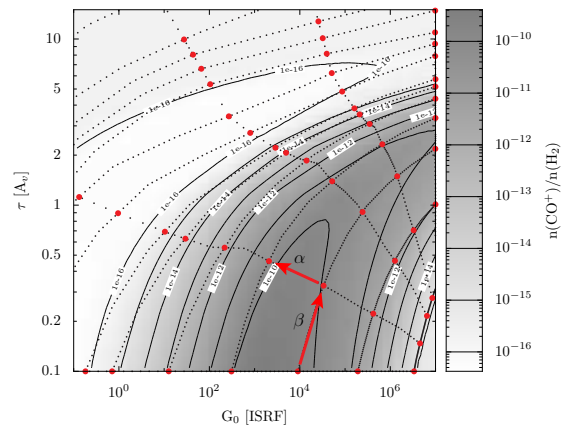}
\caption{Fractional abundance of \coplus for different FUV fluxes ($G_0$) and FUV attenuation ($\tau$). The solid contour lines and grey scale give the fractional abundance of the molecule. The gas density was chosen to be $10^6$ cm$^{-3}$, while the temperature is fixed at 550 K. The curvilinear coordinate system ($\alpha,\beta$) given in dotted line is used for the interpolation in the ($G_0,\tau$) - plane. The grid points are given by grey/red dots.\label{fig:uv_grid_co+_550k_1e6}}
\end{figure}

\citet{Staeuber04} studied the influence of FUV radiation on the chemistry in envelopes of YSOs and found \coplus to be a good tracer in high temperature regions. Observations of that molecule by \cite{Staeuber07} revealed a surprisingly high abundance which cannot be explained in terms of their spherical models. The second part of this series of papers will thus present a detailed chemical model in 2D using the interpolation approach introduced in this work. The dependence of the fractional abundance of \coplus on the radiation field $G_0$ and the attenuation $\tau$ are presented in Fig. \ref{fig:uv_grid_co+_550k_1e6}. The fractional abundance peaks at radiation strengths of $10^3$ to $10^4$ times the ISRF for low extinction ($\tau \approx 0.1$). Self-shielding can lead to a high \hh abundance at extinctions $\tau < 0.1$. We restrict the range of our chemical grid to values $\tau > 0.1$, since for the planned applications, the uncertainty in the geometry can easily account for errors of $\tau$ on the order of 0.1. At high extinction, with $\tau > 20$, the influence of the FUV radiation is negligible even for field strengths of $10^7$ times the ISRF. In absence of any attenuation, this FUV field corresponds to an early B star with a temperature of 30\,000 K and a luminosity of \tto{2}{4}~ $L_\odot$ at a distance of about 1\,000 AU.\\

%
%

An implementation of the chemical grid using $\tau$ and $G_0$ as interpolation axes would lack sufficient interpolation accuracy. Another approach was therefore developed: Contour plots of FUV sensitive molecules (e.g. C$^+$, C, CO and hydrocarbons C$_x$H$_y$) similar to Fig. \ref{fig:uv_grid_co+_550k_1e6} are calculated for different temperature regimes. The contour lines are then used to fit a curvilinear coordinate system as indicated by (red/grey) dots in Fig. \ref{fig:uv_grid_co+_550k_1e6}. A unique function relates the physical units of $\tau$ and $G_0$ to arbitrary units of a coordinate system denoted by $\alpha$ and $\beta$. In this way, the interpolation quality can be greatly improved while the number of grid points is kept constant (Table \ref{tab:param_grid}). One grid point is placed at very high extinction and no FUV irradiation to represent a model without any influence of FUV irradiation.\\

%
%
\subsection{X-ray driven chemistry} \label{sec:grid_xdriven}

Rates for the direct photoionization by X-ray photons can be evaluated in a similar way as for FUV radiation using Eq. (\ref{eq:uv_rate_integral}). The local intensity of the X-ray emission of a thermal plasma is approximated by 
\begin{equation} \label{eq:x-flux}
J(E,r) \approx \frac{L_{\rm X}}{4 \pi r^2} \cdot \frac{\mathcal{N}}{E} \cdot \exp\left(- \frac{E}{k T_{\rm X}}\right) \cdot \exp\left( -\sigma_{\rm photo}(E) \cdot N({\rm H}_{\rm tot}) \right)\ \ ,
\end{equation}
where $E$ is the photon energy, $r$ the distance to the X-ray source, $T_x$ the temperature of the X-ray emitting plasma and $N({\rm H}_{\rm tot})$ the attenuating column density. The normalization factor $\mathcal{N}$ is evaluated from the luminosity of the X-ray source $L_x$ [erg s$^{-1}$] using $L_x=4 \pi r^2 \int I_{\rm unatt.}(E,r) E dE$, where $I_{\rm unatt.}$ denotes Eq. (\ref{eq:x-flux}) without attenuation ($N({\rm H}_{\rm tot}) \equiv 0$). The photoionization cross-section $\sigma_{\rm photo}$ is obtained by summing up the contributions of all species including heavy elements in the solid phase \citep{Staeuber05}. For a photon energy above 10 keV, inelastic compton scattering (described by $\sigma_{\rm compton}$) governs the total X-ray cross-section ($\sigma_{\rm tot} = \sigma_{\rm photo} + \sigma_{\rm compton}$). The influence of elastic compton scattering and line emission in the X-ray spectra to the chemical composition can be neglected (\citealt{Staeuber05}).\\

Unlike FUV radiation, secondary processes are more important for the chemistry than direct photoionization (e.g. \citealt{Maloney96}, \citealt{Staeuber05}): Fast photoelectrons and Auger electrons can ionize other species very efficiently. The rate of ionization per hydrogen molecule due to the impact of secondary electrons can be written as
\begin{equation} \label{eq:ion_zeta_h2}
\zeta_{{\rm H}_2} = \int_{E_{\rm min}}^{E_{\rm max}} J(E,r) \sigma_{\rm total}(E) \frac{E}{W(E) x({\rm H}_2)} dE \ \ ,
\end{equation}
where $x({\rm H}_2) \approx 0.5$ denotes the fractional abundance of \hh with respect to the total hydrogen density. The mean energy per ion pair is given by $W(E)$. Like cosmic rays, secondary electrons can excite H$_2$, hydrogen and helium atoms. The electronically excited states decay back to the ground state by emitting FUV photons. Mostly photons of the Lyman-Werner bands of H$_2$ contribute to this internally generated FUV field. This field can also photodissociate and photoionize other species and hence influence the chemistry.\\

In a straight-forward implementation, three different parameters are required to parameterize the impact of X-rays on the chemistry: ({\it i}) The X-ray flux, as defined by $F_x \equiv L_x / 4 \pi r^2$, with the X-ray luminosity $L_x$ and the distance to the X-ray source $r$. ({\it ii}) The attenuating column density $N({\rm H}_{\rm tot})$ and ({\it iii}) the temperature $T_x$ of the X-ray emitting plasma. The relevant range for each grid dimension is given in Table \ref{tab:param_grid}. It is chosen to cover the range of high-mass and low-mass (class 0/1) sources as modeled by \citet{Staeuber05,Staeuber06}. The required number of 14 grid points in the dimension of the X-ray flux can be explained referring to the example of the very strong dependence of the water fractional abundance on this parameter. In Fig. \ref{fig:plot_comp_xray_zeta}, the \hho fractional abundance for a gas temperature of 120 K is shown: At an X-ray flux of $10^{-4}$ erg s$^{-1}$ cm$^{-2}$, the abundance drops by about two orders of magnitude. As discussed by \citet{Staeuber06}, this drop depends on X-ray irradiation and chemical age. Thus many grid points along the dimension of $F_X$ are required to sufficiently sample this drop.\\

Together with five points along the dimension for the cosmic-ray ionization rate $\zeta_{\rm cr}$ [s$^{-1}$], we would obtain $1.7 \times 10^7$ grid points for a straight-forward implementation (Table \ref{tab:param_grid}). This large number can be reduced by a factor of 100 using the following approximation. Instead of the X-ray model described in \citet{Staeuber05}, we make use of the fact that direct photoionization processes due to X-rays can be neglected for the chemical abundances: Instead of the X-ray model, the \hh ionization rate $\zeta_{{\rm H}_2}$ is calculated by Eq. (\ref{eq:ion_zeta_h2}) and the parameter for the cosmic-ray ionization rate is increased by this rate. A virtually enhanced cosmic-ray ionization rate thus acts as a proxy for the total ionization rate:
\begin{equation} \label{eq:totionrate}
\zeta_{\rm tot} \equiv \zeta_{\rm cr} = \zeta^0_{\rm cr} + \zeta_{{\rm H}_2}(F_x,N({\rm H}),T_x) 
\end{equation}
The ``standard'' cosmic-ray ionization rate $\zeta^0_{\rm cr} \approx 5.6 \times 10^{-17}$ s$^{-1}$ and the ionization rate of \hh due to the impact of secondary electrons $\zeta_{{\rm H}_2}(F_x,N({\rm H}),T_x)$ form together the total ionization rate, which enters the chemical model as effectively increased cosmic-ray ionization rate. In this vein, \citet{Doty04} used an increased cosmic-ray ionization rate as a proxy for an additional, internal source of ionization. Our approach treats the photodissociation reactions due to internally produced FUV photons exactly in the same way as the X-ray model by \citet{Staeuber05}: Rates by \citet{Gredel89} for cosmic ray induced FUV reactions are implemented therein with the same total ionization rate.\\

This approach provides a simple recipe to include X-ray induced reactions with an arbitrary X-ray spectrum into chemical models: The average energy per ion pair $W(E)$ and X-ray cross-sections $\sigma_{\rm photo}$ and $\sigma_{\rm compton}$ are given in Appendix \ref{sec:app_xcross}. Using Eq. (\ref{eq:ion_zeta_h2}) the total ionization rate $\zeta_{{\rm H}_2}(F_x,N({\rm H} ),T_x)$ can be calculated. Pre-calculated ionization rates for a thermal X-ray spectrum (Eq. \ref{eq:x-flux}) are given in the same appendix.\\

%
%
\subsection{Total ionization rate vs. X-rays} \label{sec:ionratxray}

\begin{figure}[tbh]
\plotone{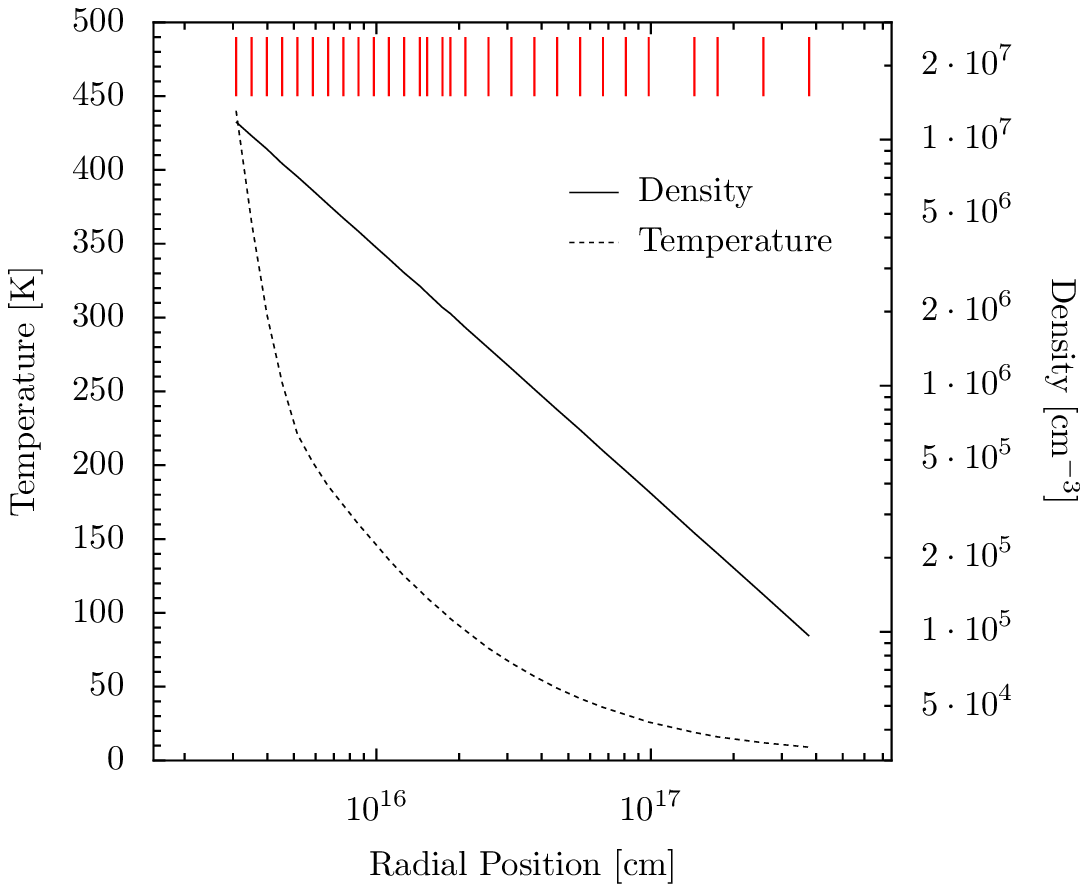}
\caption{Density and temperature in the spherical model of AFGL 2591. Vertical lines on top of the figure indicate the position of the shells used for the calculation.\label{fig:struc_afgl}}
\end{figure}

The approach to use a total ionization rate instead of the previous X-ray model is tested with a spherical model of AFGL 2591 as given in \citet{Staeuber05}. In this model, temperature and density vary with distance to the central source. The density structure is given in the form of a power-law in radius (\citealt{vdTak99}) and the temperature structure is taken from the detailed thermal balance calculations of \citet{Doty02}. The temperature and density distribution are given in Fig. \ref{fig:struc_afgl}. The model is devided into 30 shells of approximately constant column density in radial direction. In this way, more shells are placed in the warm, chemically active part of the envelope. To resolve the water evaporation properly, assumed to be at 100 K, an additional shell is set to the corresponding position. Fewer positions are needed in the outer, colder and thus chemically less active region. Calculations with more shells show insignificant differences in observable quantities derived from the models. More points may be necessary for other temperature/density profiles as presented in Fig. \ref{fig:struc_afgl}. When the interpolation method for chemical abundances is used to construct such a spherical model, the number of points can be easily increased due to the gain in speed using that approach. For the implementation, we suggest to bisect the points until the derived observable quantities converge.

%
%

\begin{figure*}[tbh]
\includegraphics[width=\textwidth]{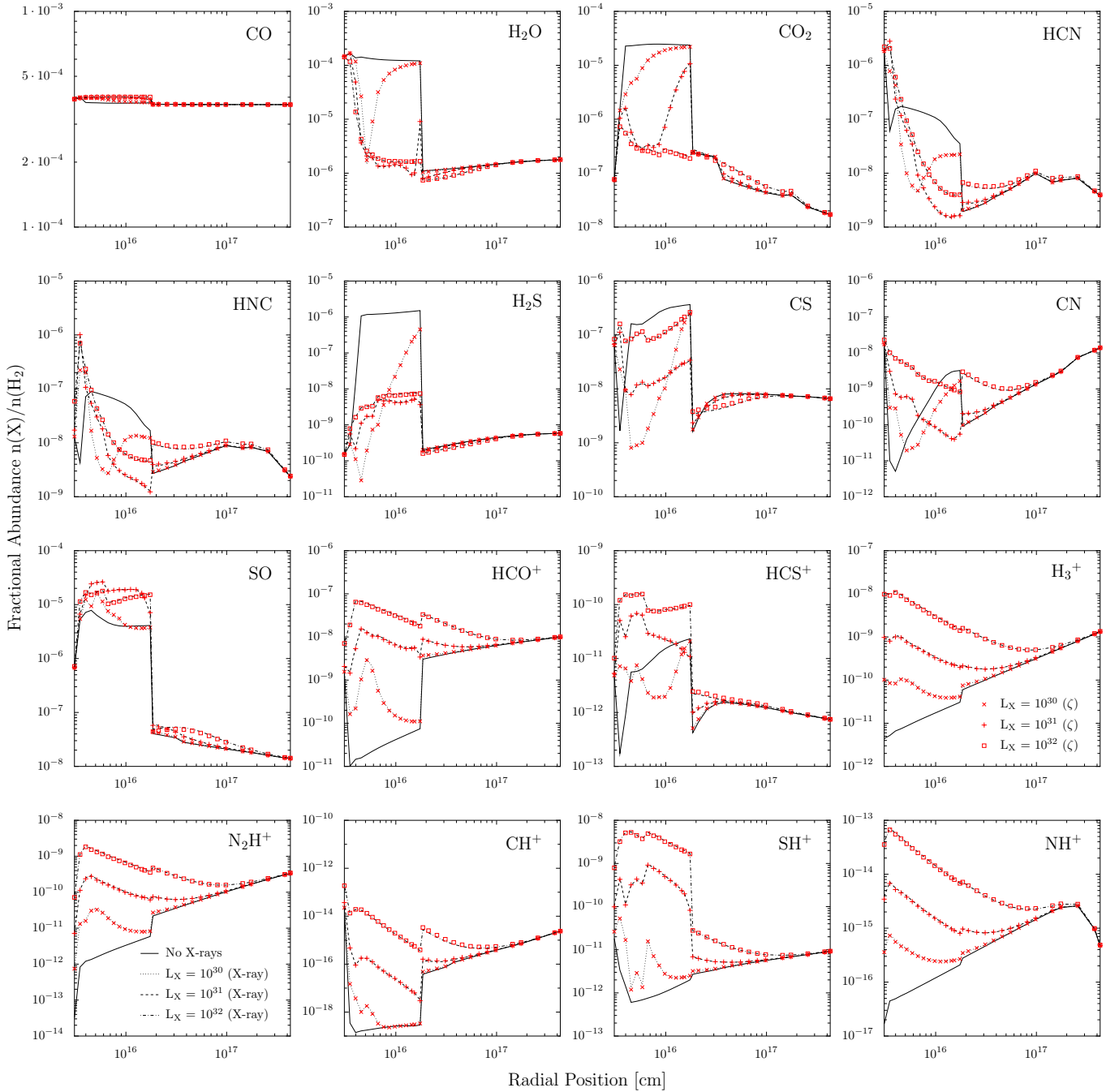}
\caption{Abundances in a spherical AFGL 2591 model calculated by the X-ray model of St\"auber et al. (``X-ray'', black lines) compared to a model with a total ionization rate given by Eq. (\ref{eq:totionrate}) (``$\zeta$'', grey/red lines). Four different models are shown for X-ray luminosities of $0$, $10^{30}$, $10^{31}$ and $10^{32}$ erg s$^{-1}$.\label{fig:compare_afgl_radial_zeta}}
\end{figure*}

Figure \ref{fig:compare_afgl_radial_zeta} shows the radial dependence of the abundances for the main molecules CO, H$_2$O, CO$_2$ as well as for those 10 molecules predicted to be the best X-ray tracers by \citet{Staeuber05} (HCN, HNC, H$_2$S, CS, CN, SO, HCO$^+$, HCS$^+$, H$_3^+$ and N$_2$H$^+$). The ionized hydrides CH$^+$, SH$^+$ and NH$^+$, discussed later in this section, are given in addition. The X-ray models are indicated by black lines. Results from the model with a total ionization rate $\zeta$ given by Eq. (\ref{eq:totionrate}) are presented by thick (grey/red) lines. A model without protostellar X-ray radiation and models with X-ray luminosity $L_X$ = $10^{30}$, $10^{31}$ and $10^{32}$ erg s$^{-1}$ are shown. Protostellar FUV radiation is included in the same way as \citet{Staeuber05} assuming a flux of $G_0$ = 10 at 200 AU. The agreement between the model including X-rays and with a total ionization rate is very good for the shown set of species, with deviations less than $25\%$ in the fractional abundance.\\

%
%

This very good agreement between X-ray models and models with a total ionization rate has observational consequences: In order to use molecular lines as tracers for protostellar X-ray radiation, the effects of cosmic-ray ionization and X-ray induced ionizations have to be disentangled by spatial or excitation information on the abundance. This can be obtained from high-$J$ lines with a high critical density (e.g. observed by the upcoming HIFI spectrometer onboard the Herschel Space Observatory) or by comparing visibility amplitudes from high angular resolution interferometer observations (e.g. \citealt{Benz07}).\\

%
%
\begin{figure*}[tbh]
\includegraphics[width=\textwidth]{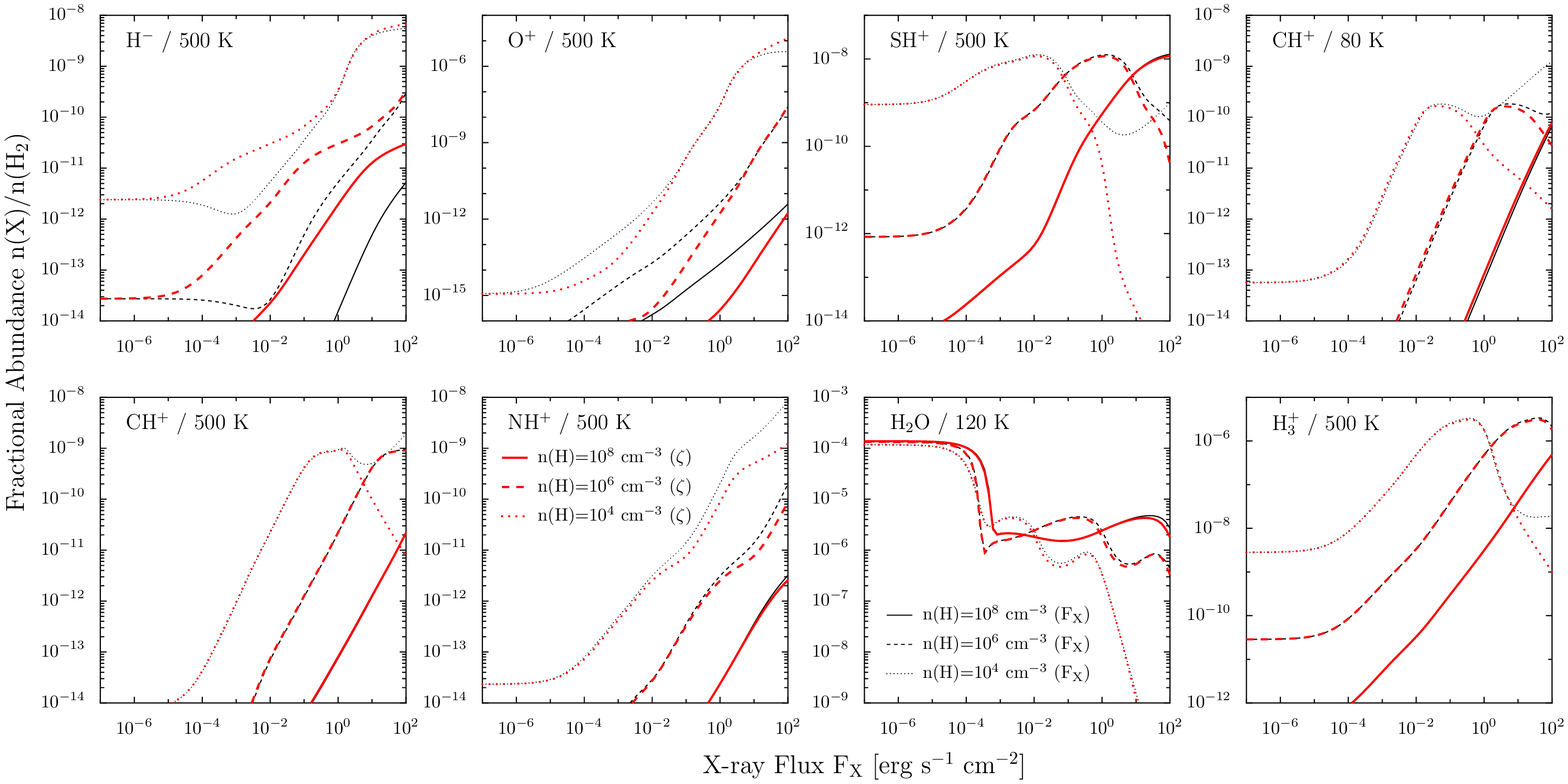}
\caption{Abundance versus X-ray luminosity of species with significant deviations between the X-ray model (\textit{black lines}) compared to the results of the model with enhanced cosmic-ray ionization rate (\textit{grey/red lines}). The temperature is indicated on the top. Models for different total hydrogen densities ($10^4$, $10^6$ and $10^8$ cm$^{-3}$) are given to point out the largest effects. Water exhibiting little deviation is shown as an example for comparison.\label{fig:plot_comp_xray_zeta}}
\end{figure*}

What are the limits of this approach? The X-ray flux $F_X = L_X / 4 \pi r^2$ at a distance of 20 AU from a source with an X-ray luminosity of $10^{32}$~ erg s$^{-1}$ is about 90 erg s$^{-1}$ cm$^{-2}$. \citet{Meijerink05a} use similar field strengths when modeling X-ray dominated regions (XDRs) in galactic nuclei. Thus we explored the validity of the ionization rate approximation up to X-ray fluxes of 100 erg s$^{-1}$ cm$^{-2}$. All molecules in the chemical network have been tested for significant deviations between the two approaches in the range from 0 to 100 erg s$^{-1}$ cm$^{-2}$. Excellent agreement as in the example of water (Fig. \ref{fig:plot_comp_xray_zeta}g) was found for most species. For our application in envelopes of YSOs, the ionization rate approach thus can be safely used.\\

A small set of species revealed discrepancies between the two approaches under certain conditions. They are presented in Fig. \ref{fig:plot_comp_xray_zeta}. There are different reasons for deviation. They are discussed in the following paragraphs.\\

The negatively charged species H$^-$, OH$^-$ and CN$^-$ significantly deviate since the X-ray approach does not include a path for the formation of H$^-$ by the ionization of H$_2$ (Fig. \ref{fig:plot_comp_xray_zeta}a). The UMIST database lists the reaction H$_2$ + c.r. $\rightarrow$ H$^+$ + H$^-$ for this process, resulting in a much higher abundance of this species. Subsequently, the abundances of OH$^-$ and CN$^-$ also increase. The X-ray approach only includes H$^-$ formation by radiative association (H + $e^-$ $\rightarrow$ H$^-$ + $\gamma$). Thus our total ionization rate approach to is an improvement.\\

At low X-ray fluxes the abundance of O$^+$ is higher in the X-ray model than in the total ionization rate approach (Fig. \ref{fig:plot_comp_xray_zeta}b). In the X-ray approach, direct photoionization CO $+$ $\gamma_{\rm X}$ $\rightarrow$ C$^+$ + O$^+$ + 2$e^-$~ can compete with the relatively inefficient formation of O$^+$ through the reaction of He$^+$ with CO$_2$, OCS or SO. The fractional abundance of O$^+$ thus increases due to the presence X-ray photons. However, it remains below a few times $10^{-12}$, insignificant for our applications (Sect. 4.1). At higher X-ray fluxes, O$^+$ is formed in the charge exchange reaction of H$^+$ with O, and the two approaches agree well.\\

For the hydride ions SH$^+$ and CH$^+$ at low density and high X-ray flux, the X-ray approach predicts higher abundances than the ionization rate approach. This discrepancy can be explained by doubly ionized atoms added by \citet{Staeuber05} to the chemical network, but not included in the UMIST database. Thus doubly ionized atoms lack in the total ionization rate approach. At high densities, doubly ionized atoms are not important because X + \hhhplus $\rightarrow$ XH$^+$ + \hh dominates XH$^+$ formation. \citet{Abel08} found the branching ratio of S$^{++}$ + H$_2$ to be important for the efficiency of SH$^+$ production through doubly ionized atoms. \citet{Staeuber05} only included X$^{++}$ + H$_2$ $\rightarrow$ X$^+$ + H$_2^+$ with rate coefficients taken from \citet{Yan97} ($k$ $\leq$ 10$^{-13}$~ cm$^3$ s$^{-1}$). However, in the later version of their code used here, They implemented the formation of hydride ions as the main product with a rate coefficient of 10$^{-9}$~ cm$^3$ s$^{-1}$ (\citealt{Yan97}), similar to the value of \citet{Abel08} for SH$^+$, but about an order of magnitude higher for CH$^+$.\\

Doubly ionized atoms become important for low densities and high irradiation. For a given X-ray flux, the number of ionization processes per volume is constant, independent of the total gas density. The electron density at a high X-ray flux of $10^2$~ erg cm$^{-2}$ s$^{-1}$ varies only by a factor of about two between a density of $10^4$ and $10^8$~ cm$^{-3}$. The amount of O$^+$ is large, accounting for a short destruction time-scale of \hh in the reaction \hh + O$^+$ $\rightarrow$ \ohplus + H. \hh is however the precursor of \hhhplus which is subsequently also reduced (Fig. \ref{fig:plot_comp_xray_zeta}h). In this regime, the main production of SH$^+$ and CH$^+$ is through X$^{++}$ + H$_2$ $\rightarrow$ XH$^+$ + H$^+$. \\

NH$^+$ enhancement in the X-ray approach is due to another effect. The charge exchange reactions of oxygen and nitrogen with H$_2$ have rate coefficients of \tto{3}{-11} cm$^3$ s$^{-1}$ and $10^{-9}$ cm$^3$ s$^{-1}$, much faster than for sulphur and carbon. The chemical network of \citet{Staeuber05} follows \citet{Yan97} and thus assumes the formation of OH$^+$ and NH$^+$ through doubly ionized species to be negligible. The slightly enhanced abundance of NH$^+$ in the X-ray approach compared to the total ionization rate approximation is explained by the larger amount of N$^+$ at low density, since its production by the reaction of N$^{++}$ + H $\rightarrow$ H$^+$ + N$^+$ is faster than the reaction between N$^{++}$ and H$_2$. \\

In conclusion, the total ionization approach should not be used for low densities where the X-ray fluxes exceeds 10 erg s$^{-1}$ X-ray flux. For SH$^+$ and CH$^+$, the X-ray flux limit is lower (Fig. \ref{fig:plot_comp_xray_zeta}). We note that the X-ray approach is also incomplete in this regime as e.g. the important vibrationally excited H$_2$ with a high energy deposition of X-rays per density $H_X/n > 10^{-25}$~ erg cm$^3$ s$^{-1}$ (\citealt{Yan97}) is not included. Nevertheless, the total ionization rate approach is well applicable for models of YSO envelopes, since density models by \citet{Jorgensen02} or \citet{vdTak00a} predict a gas density larger than 10$^6$ cm$^{-3}$ close to the protostar where X-ray irradiation is significant (i.e. exceeding than the cosmic ray effects). Indeed, the derived abundances of SH$^+$, CH$^+$ and NH$^+$ in Fig. \ref{fig:compare_afgl_radial_zeta}, show no difference between the two approaches. For our application, the total ionization approach is not only more elegant, but adequate.

%
%

\subsection{Multidimensional interpolation} \label{sec:multidimint}

The relevant physical parameters for the molecular/atomic fractional abundance evolution are the temperature $T$, the density $n$, two parameters for the FUV flux and the total ionization rate. The implemented grid thus consists of five dimensions and a total of about \tto{1.1}{5} grid points in the improved implementation. The range and number of points for each dimension are given in Table \ref{tab:param_grid}. For the interpolation of the abundance at a specific physical condition $\vec{\lambda} \equiv \left( \lambda^1,\lambda^2,\lambda^3,\lambda^4,\lambda^5 \right)$ a multidimensional interpolation in logarithmic space is used: The neighboring interpolation points of $\vec{\lambda}$ are found in each dimension $d=\left\lbrace 1,2,\ldots,5 \right\rbrace$, such that $\lambda^d_{\rm a} < \lambda^d < \lambda^d_{\rm b}$. Then, the abundances at the interpolation points, $x\left(\lambda^1_i,\lambda^2_j,\lambda^3_k,\lambda^4_l,\lambda^5_m \right)$ are read out of the grid for all combinations of $i,j,k,l,m=\lbrace {\rm a,b} \rbrace$. The interpolated abundance $x$ is obtained from 
\begin{eqnarray*}
\log_{10}(x) = \sum_{i,j,k,l,m=\lbrace {\rm a,b} \rbrace}&{}& \alpha^1_i \, \alpha^2_j \, \alpha^3_k \, \alpha^4_l \, \alpha^5_m \\
&{}&\cdot \log_{10}\left(x\left(\lambda^1_i,\lambda^2_j,\lambda^3_k,\lambda^4_l,\lambda^5_m \right)\right) \ .
\end{eqnarray*}

The weights $\alpha^d_i$ along each dimension are defined by $\alpha^d_a=1-\beta$ and $\alpha^d_b=\beta$. The position within the hypercube, normalized to the interval between 0 and 1 is given by $\beta=\left[ \log_{10}(\lambda^d)-\log_{10}(\lambda^d_{a}) \right]/\left[\log_{10}(\lambda^d_{b})-\log_{10}(\lambda^d_{a})\right]$.\\

The most time-consuming step of the interpolation is the calculation of the total ionization rate by the integral of Eq. (\ref{eq:ion_zeta_h2}) and the inversion of the coordinate transformation $(\alpha,\beta) \rightarrow (G_0,\tau)$. Both functions are given in tabulated form in the interpolation routine. This allows to obtain approximately 10 000 fractional abundances per second on a standard personal computer. Using the full chemical model, the calculation of the same number of abundance evolutions would require about 80 hours of CPU time.

%
%
\section{Chemical model} \label{sec:chemmod}

For our work, we start from the chemical model introduced by \citet{Doty02,Doty04} and \citet{Staeuber04,Staeuber05}. Thus we describe only changes to their model in this section: The chemical network has been updated from the UMIST 97 (\citealt{Millar97}) to the UMIST 06 database for astrochemistry (\citealt{Woodall07}). We implement the standard UMIST 06 database without the dipole-enhanced rates. For technical reasons, fluorine bearing species have not been included. Appendix \ref{sec:app_umist} gives details of the implementation of the new chemical network. The rates for cosmic ray and X-ray induced FUV reactions implemented in \citet{Staeuber05} were a factor of 2 too high due to an error in the previously used database (cf. comment by S. Doty in \citealt{Woodall07}) and are now at their correct values. Charge exchange on PAH or small grains can be important for the ionization balance (\citealt{Wakelam08}, \citealt{Maloney96}). FUV irradiation enhances the number of positively charged grains. The rates for this process were accidentally divided by the total density in the implementation of \citet{Staeuber05} and are now at the correct value. The influence on the chemical abundance in their applications is however small. Grain surface reactions are not taken into account except for the formation of \hh, where the rate given in \citet{Draine96} is used.\\

We have extended the chemical model with self-shielding of CO and \hh using the shielding factors given in \citet{Lee96} and \citet{Draine96}, respectively. They depend on the column densities $N({\rm H}_2)$ and $N({\rm CO})$ between the FUV source and the modeled parcel of gas. Since we do not include any ``non-local'' parameter as dimension in the chemical grid, an approximation is needed: Prior to the calculation of a model, the optical depth-dependence of the \hh abundance at the given physical conditions (density, temperature and FUV irradiation) is calculated for a simple steady state-model considering only the \hh photodissociation and formation on dust with a fixed density and temperature. The column density can be read out of this toy model and shows good agreement with the examples given in \citet{Draine96}. As a rough approximation of the CO column density, we assume a fixed ratio of $N({\rm CO})/N({\rm H}_2)=2 \times 10^{-4}$.\\

In the papers of Doty et al./St\"auber et al., the system of stiff ordinary differential equations (Eq. \ref{eq:rate}) is solved using the \verb!DDRIV3! algorithm\footnote{\anchor{http://www.netlib.org/slatec}{http://www.netlib.org/slatec}}. This solver is numerically unstable for chemical networks involving reactions with short timescales, e.g. evaporation from dust or photodissociation at high values of G$_0$. The \verb!DVODE! solver\footnote{\anchor{http://www.netlib.org/ode}{http://www.netlib.org/ode}} proved to be more robust and much faster in a large range of physical parameters. Similar characteristics in comparing the two algorithms were found by \citet{Nejad05}. The equations for the conservation of elements revealed a better accuracy of the \verb!DVODE! solver. A single run of the chemical model on a standard personal computer takes about 30 s, and the calculation of the whole grid thus about 900 hours of CPU time. Distribution on several CPUs is easily possible and allows to build various chemical grids e.g. for changed chemical networks or different initial conditions in relatively short time.

%
%
\subsection{Initial conditions} \label{sec:ics}

In order to solve the first order differential rate equations (Eqs. \ref{eq:rate}), initial conditions for the abundances have to be assumed. If the chemical evolution were traced starting from diffuse cloud conditions, a purely atomic composition would have to be assumed and the physical conditions such as FUV irradiation or density would have to be changed during the evolution (e.g. \citealt{Lintott06}). Here we follow the approach of Doty/St\"auber et al., who started at dark cloud conditions with a molecular composition (Table \ref{tab:init_cond} in the appendix). In this way, uncertainties of the physical evolution during this first phase and/or reactions on dust grains less affect the chemical composition. The evaporation of species frozen out on ice is not taken explicitly into account with evaporation-type reactions but approximated with different sets of initial abundances as in the models of Doty/St\"auber et al. Evaporation and freeze-out reactions are implemented in the model. However they slow down the calculation significantly due to short timescales at high temperature and are not activated for this work.\\

\subsection{Sulphur bearing species} \label{sec:ics_sulphur}

Various sulphur bearing molecules are predicted by \citet{Staeuber05} to be good tracers of X-ray radiation. Despite the large uncertainty in many of the reaction rates involving sulphur bearing species (\citealt{Wakelam04a}), these molecules might be important for applications of the grid. To better constrain the initial molecular conditions, we use the multitude of observed sulphur bearing species found toward AFGL 2591 by \citet{vdTak03}, \citet{vdTak99} and \citet{Staeuber07}. Spherical models of AFGL 2591 (cf. Sect. \ref{sec:ionratxray}) are calculated for different sets of initial abundances, main sulphur carriers and chemical ages. The abundances are used as input for a full non-LTE radiative transfer calculation using the RATRAN code (\citealt{Hogherheijde00}) to model line fluxes. A $\chi^2$ test is carried out to compare the modeled line fluxes with the observations. For this work we adopt S ($T<100$ K) and \hhs ($T>100$K) to be the main initial sulphur carriers. The abundance for the cold part follows \citet{Aikawa08} to be \tto{9.1}{-8} relative to the total hydrogen density. For the hot core, the abundance of SO is chosen to reproduce the jump in the abundance between the cold and hot part as observed by \citet{Benz07}. The adopted sulphur abundance is about the solar abundance (cf. \citealt{Snow96} and \citealt{Asplund05}) indicating no or only minor sulpur depletion on dust grains in the hot core. \citet{Goicoechea06} modeled a PDR with a relatively low FUV irradiation of $\chi=60$ ISRF (Draine) and reported a sulphur abundance of about a factor of 4 lower than the value adopted in this work to reproduce their observations. It may be caused by the massive impact of FUV irradiation in the vicinity of AFGL 2591, raising the temperature severely and resulting in an even higher sulphur abundance in the gas phase.

%
%
\section{Benchmarks} \label{sec:benchmark}

In this section, the accuracy of the interpolation method is verified. For two realistic problems -- possible applications of the chemical grid -- the interpolated abundances are compared to abundances calculated using the chemical model. The results are considered satisfactory if they agree within a factor of two of each other for the reasons noted in Sect. \ref{sec:grid}.

%
%
\subsection{A spherical model of AFGL 2591} \label{sec:comp_afgl}

\begin{figure*}[ht]
\includegraphics[width=0.90\textwidth]{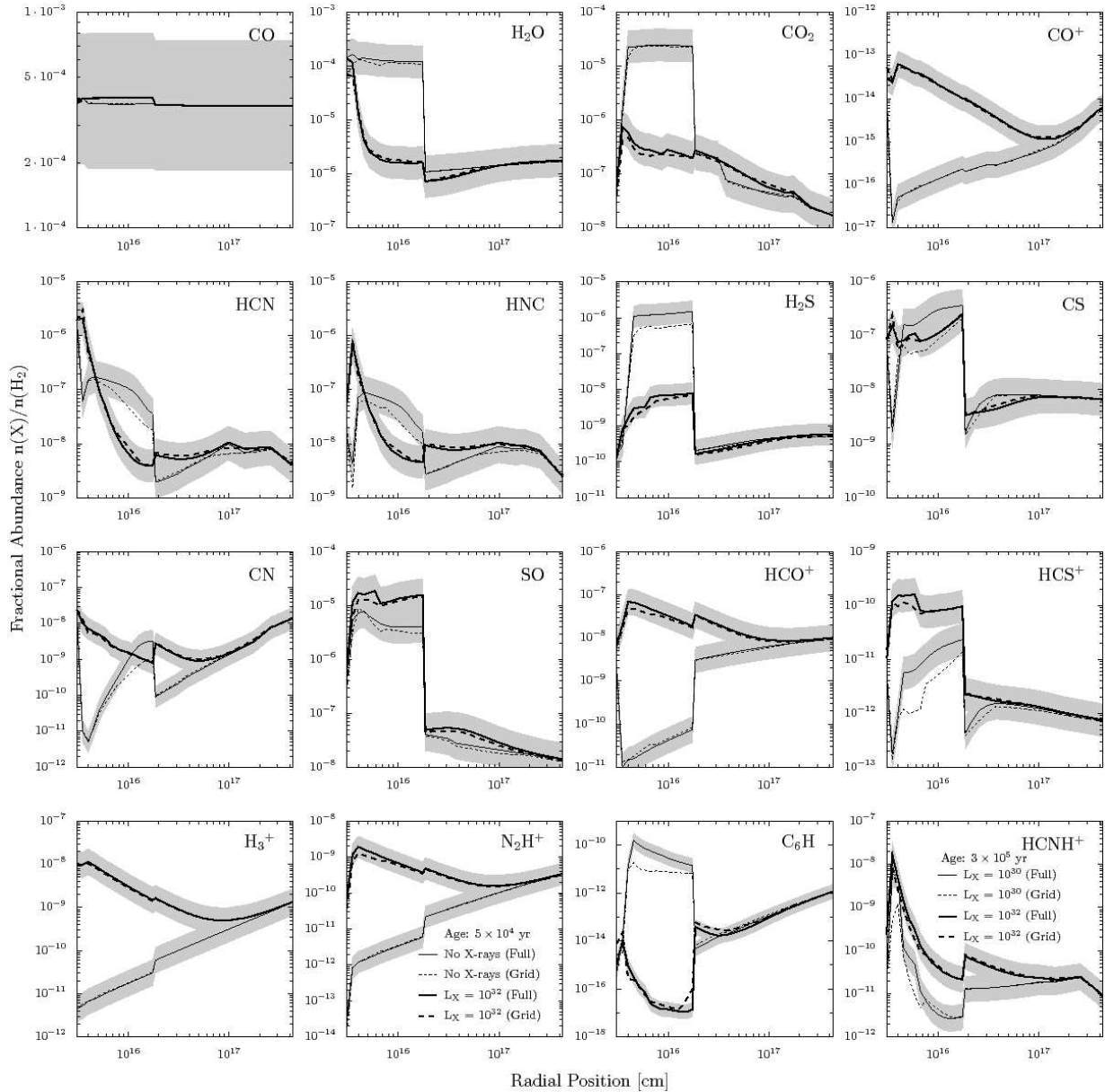}
\caption{Abundances of major species, X-ray/UV tracers and problem cases in a spherical model of AFGL 2591. The plots show a comparison between results calculated by the chemical model (solid line) and interpolated from the grid of chemical models (dashed line). The grey region shows a deviation of a factor of 2 from the chemical model. Note the different vertical scales of individual panels. The thick lines corresponds to a model with a stellar X-ray luminosity of $10^{32}$ erg s$^{-1}$ and the thin line to a model without X-rays. All abundances are given for a chemical age of \tto{5}{4} yr, except for HCNH$^+$ (\tto{3}{5} yr).\label{fig:compare_afgl_radial}}
\end{figure*}

A first test of the chemical grid is performed using the spherical model of AFGL 2591 as introduced in Sect. \ref{sec:ionratxray}. Figure \ref{fig:compare_afgl_radial} shows the radial dependence of the abundances for the main molecules CO, H$_2$O, CO$_2$ as well as for those 10 molecules predicted to be the best X-ray tracers by \citet{Staeuber05} (HCN, HNC, H$_2$S, CS, CN, SO, HCO$^+$, HCS$^+$, H$_3^+$ and N$_2$H$^+$). \coplus which will be addressed in the next paper of this series and the two molecules, C$_6$H and HCNH$^+$, having the largest deviation are shown in addition. Solid lines give the results of the chemical model, while interpolated abundances are shown by dashed lines. Models with no protostellar X-ray radiation and an X-ray luminosity of $L_{\rm X}=10^{32}$ erg s$^{-1}$ (\citealt{Staeuber05}) are given in thin and thick lines, respectively. We assume a chemical age of \tto{5}{4} yr. The shaded region indicates a range of a factor of two compared to the fully calculated model and marks the goal of our interpolation approach. Indeed, most molecules comply with the aimed accuracy for the models with and without X-rays.\\

To provide an unbiased check for a larger set of molecules, a statistical approach is used. The goal of chemical modeling is to obtain characteristics which can be compared to observations. In the following, we check the column density of all molecules in the chemical model for agreement between grid and full calculation. The radial column density, $N_{\rm radial} = \int n(r) dr$, with the radial distance $r$, gives a measure for molecules observed in absorption. For molecules observed in emission, the beam averaged column density is more appropriate. It is defined by
\begin{equation} \label{eq:beamavg}
N_{\rm beam} = \frac{\int \int n(z,p) G(p) \, 2 \pi p \, dp \, dz}{\int G(p) \, 2 \pi p \, dp} \ ,
\end{equation}
with $z$ being the coordinate along the line of sight and $p$ the impact parameter perpendicular to $z$. $G(p)$ is a beam response function (e.g. a gaussian). This approach using column densities does not incorporate the molecular excitation. Thus it cannot be used to compare models directly with observations. Computationally much more demanding radiative transfer calculations would have to be carried out instead. For our benchmark purposes however, column densities are sufficient.\\

%
%

Four different spherical models of AFGL 2591 without protostellar X-rays and with X-ray luminosity of $L_x=10^{30}$,$10^{31}$ and $10^{32}$ erg s$^{-1}$ are compared. Three different chemical ages of \tto{3}{3}, \tto{5}{4} and \tto{3}{5} yr are considered to trace possible deviations due to the temporal evolution of the fractional abundances. For these 12 models, the radial ($N_{\rm radial}$) and beam averaged ($N_{\rm beam}$) column density are obtained for all molecules in the chemical network. We assume a 14'' beam corresponding approximately to a JCMT or Herschel beam and adopt a distance of 1 kpc to the source. To compare grid results to fully calculated models, the factorial deviation $Y \equiv \textrm{max}\left(N_{\rm grid}/N_{\rm full},N_{\rm full}/N_{\rm grid}\right)$ is introduced. The fraction of molecules having a factorial deviation larger than a certain value is shown in Fig. \ref{fig:compare_stat}. Only molecules with a column density larger than $10^{11}$~ cm$^{-2}$ are considered for the following reason: Assuming optically thin radiation, the line strengths corresponding to this column density for a molecule at maximal excitation, a typical Einstein-A coefficient of $10^{-3}$~s$^{-1}$ and a frequency of 350 GHz is about 40 mK km s$^{-1}$, a lower limit for observations with current and upcoming facilities. Compared to the H$_2$ column density of order $10^{23}$ cm$^{-2}$, it corresponds to a fractional abundance of about $10^{-12}$ which marks approximately the possible limit of a detection within an observation time of a few hours.\\

Table \ref{tab:uvbads} in Appendix \ref{sec:dissagree} lists the molecules with the largest deviations. The largest disagreement is found in the radial column density of OH$^-$ with a factorial deviation up to a factor of 9.6. This deviation is explained by the different paths for H$^-$ formation in the two models (Sect. \ref{sec:ionratxray}). It is thus an improvement of the total ionization approach. A problematic disagreement larger than a factor of 5 is found however for HCNH$^+$ ($Y=5.4$) and C$_6$H ($Y=5.2$). Figure \ref{fig:compare_afgl_radial} shows the radial dependence of the fractional abundances for these two species at the chemical age and X-ray luminosity where the largest deviation occurs.\\

\begin{figure}[ht]
\plotone{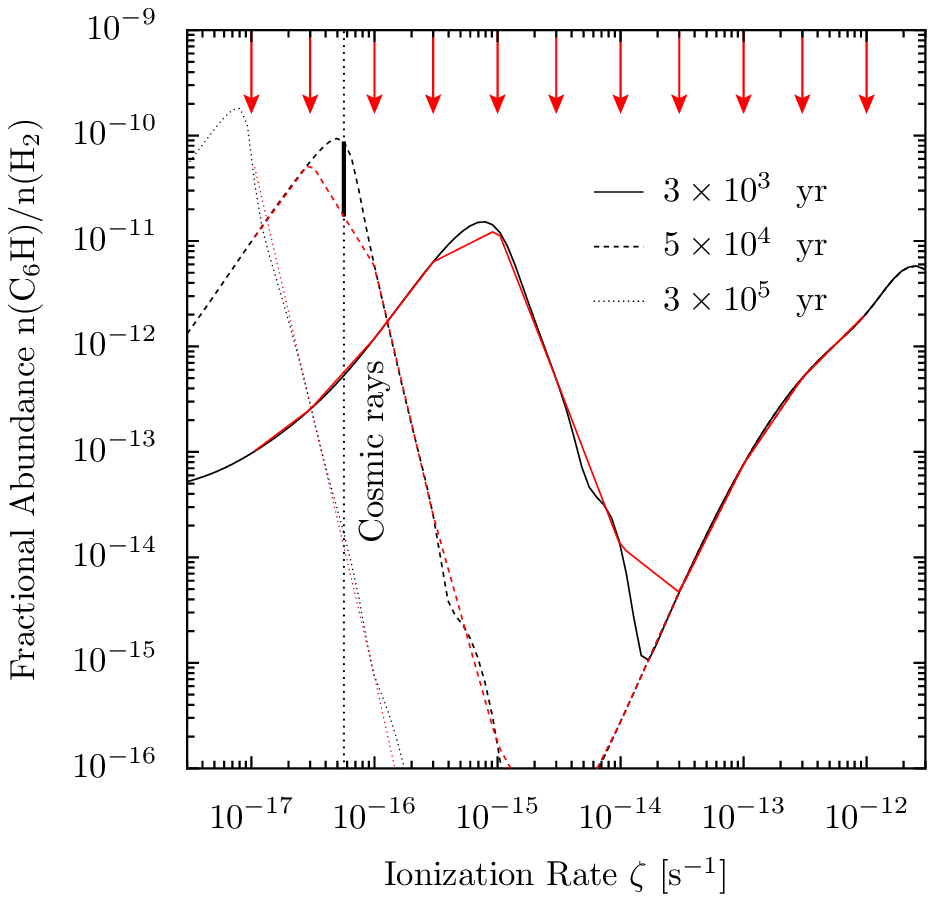}
\caption{Illustration of the C$_6$H problem: Fractional abundance of C$_6$H vs. ionization rate calculated by the full chemical model (black lines) and from the grid of chemical models (grey/red lines). Three different chemical ages are given. The grid points are indicated at the top of the figure. The cosmic-ray ionization rate of \tto{5.6}{-17} s$^{-1}$ is given by the dotted line.\label{fig:afgl_devi_a}}
\end{figure}

How can such large deviations be explained? In the case of C$_6$H at low X-ray fluxes, it is surprisingly an effect of the ionization rate. Figure \ref{fig:afgl_devi_a} presents the fractional abundance of C$_6$H vs. total ionization rate obtained from the chemical model and interpolated from the chemical grid. Three different chemical ages are given. For \tto{5}{4} yr, we find a deviation of about a factor of 5 due to the interpolation over a sharp peak in the fractional abundance (bold vertical bar), which is not resolved by the grid. This peak shifts to higher ionization rates with time. A proper sampling would thus require a large number of grid points to cover all chemical ages. At ionization rates below this peak, C$_6$H is formed by the reaction C + C$_5$H$_2$ $\rightarrow$ C$_6$H. C stems from the X-ray or cosmic ray induced dissociation of CO. At an older chemical age (or higher ionization rate), the ``late-phase'' molecule SO$_2$ is photodissociated to SO and O. Reaction of C$_6$H + O then decreases the fractional abundance of C$_6$H. At very high ionization rate, the bulk of C$_6$H is produced by the electron recombination of C$_7$H$^+$.\\

\begin{figure}[ht]
\plotone{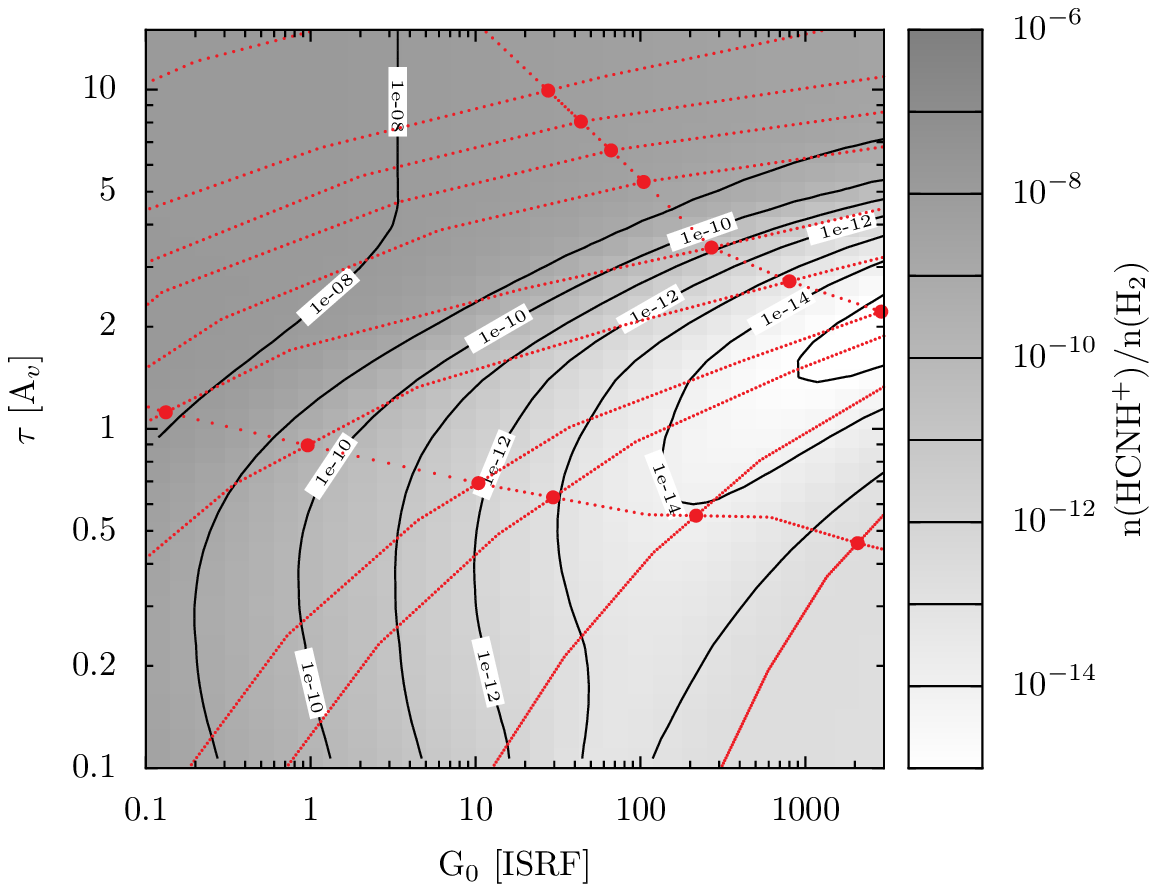}
\caption{Illustration of the HCNH$^+$ problem: Fractional abundance of HCNH$^+$ (grey scale and contours) at a chemical age of \tto{3}{5} yr and physical conditions corresponding to the innermost part of the AFGL 2591 model with $L_{\rm X}=10^{30}$ erg s$^{-1}$. The dotted grey/red lines indicate the coordinate system used for the interpolation and the solid points correspond to grid points.\label{fig:afgl_devi_b}}
\end{figure}

Most contribution to the radial column density of HCNH$^+$ is from the innermost part of the model. In this region, HCNH$^+$ is mainly formed by the reaction of HCN with H$_3$O$^+$. Both molecules are enhanced by X-rays and FUV irradiation at temperatures above 250 K, where water is not destroyed due to X-ray irradiation (\citealt{Staeuber06}). An X-ray luminosity higher than $10^{30}$ erg s$^{-1}$ is sufficient to dominate the effect of FUV enhancement. At lower X-ray luminosities however, the FUV flux dominates the HCNH$^+$ abundance. Due to the particular shape of the contour lines of its fractional abundance depending on $G_0$ and $\tau$, this molecule is difficult to interpolate. Figure \ref{fig:afgl_devi_b} shows the fractional abundance of HCNH$^+$ in the $(G_0,\tau)$-plane along with the coordinate system used for the gridding and interpolation (Sect. \ref{sec:uvdriven}). Physical conditions of the innermost region of the AFGL 2591 model and a chemical age of \tto{3}{5} yr were chosen for this figure.\\

%
%

%
%

In general, the agreement between the grid results and the fully calculated abundances is excellent. The mean deviation of $\left\langle Y_{\rm beam} \right\rangle = 1.12$ for the beam averaged column densities and $\left\langle Y_{\rm radial} \right\rangle = 1.28$ for the radial column densities indicate agreement for the majority of the species. A total number of 1661 ($N_{\rm radial}$) and 1563 ($N_{\rm beam}$) column densities are considered for the means, respectively. The slightly higher mean deviation of the radial column density is explained by the larger weight of the chemically active hot-core region. The mean deviation does not vary by more than 0.04 with chemical age.

%
%
\subsection{FUV driven chemistry} \label{sec:uvchem}

%
%
\begin{figure*}[tbh]
\includegraphics[width=\textwidth]{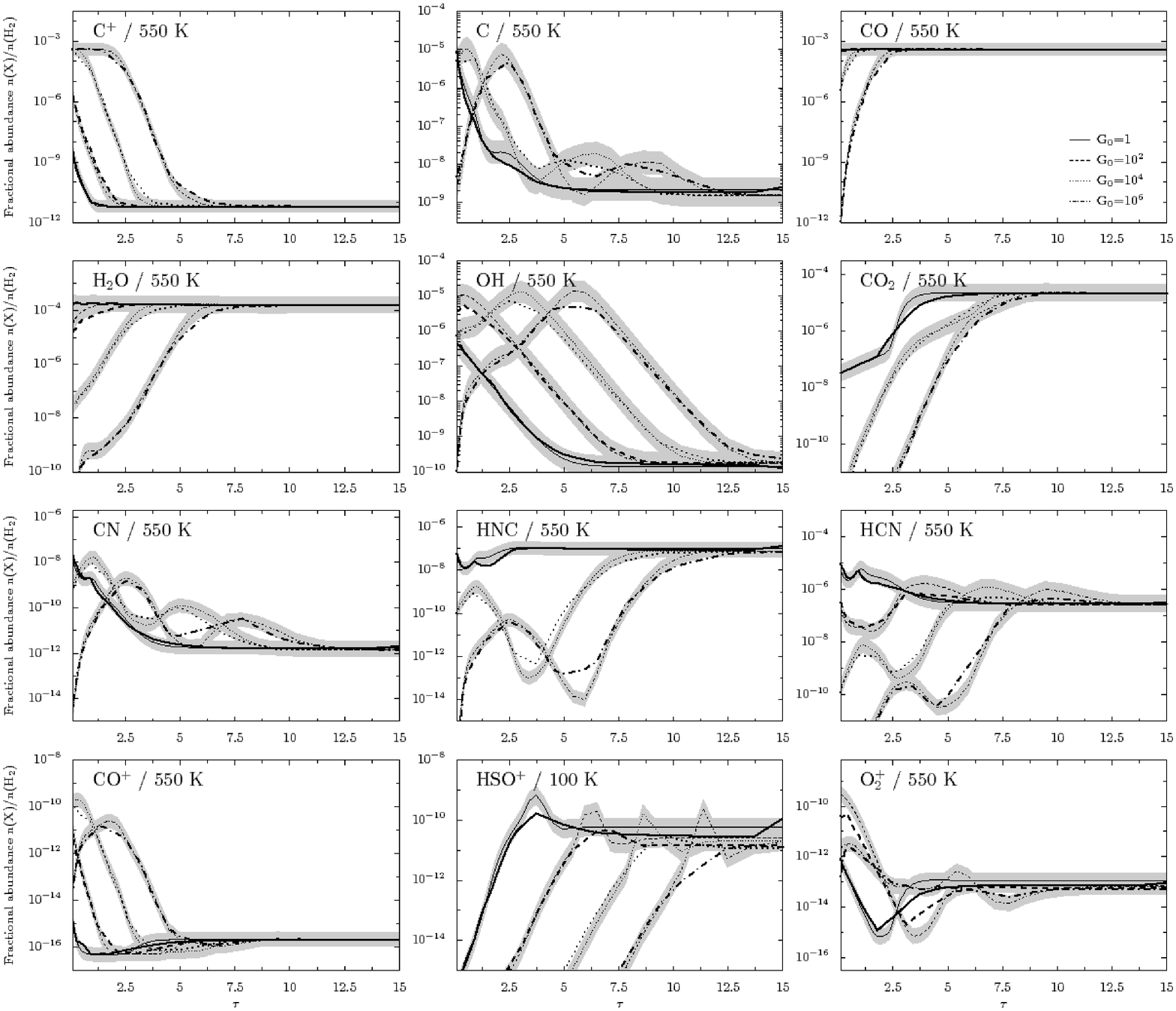}
\caption{Fractional abundances of FUV sensitive species between $\tau=0$ and $\tau=15$. For this presentation, the temperature is fixed at the given value and the density is assumed to be $10^6$ cm$^{-3}$. Results for an incident FUV field of 1, $10^2$, $10^4$ and $10^6$ ISRF are shown. The thick line corresponds to the result calculated by the chemical model while the thin line gives the results from the grid interpolation. The grey region indicates the range of a factor of two relative to the calculated model specified for agreement of the interpolation.\label{fig:plot_uv_slides}}
\end{figure*}

In the spherically symmetric models of AFGL 2591, protostellar FUV radiation cannot penetrate further into the envelope than a few hundred AUs (\citealt{Staeuber04}). This is however not the case when an outflow cavity allows FUV photons to escape and irradiate a larger volume of gas. This 2D situation will be addressed in the second and third parts of this series of papers. Here we carry out the following benchmark test. A (plane parallel) region with a fixed temperature of 80, 100 and 550 K and a density of 10$^6$~ cm$^{-3}$ is considered. The incident radiation field is assumed to be 1, 10 10$^2$, 10$^4$ and 10$^6$ times the ISRF and no geometrical dilution (i.e. $G_0 \propto r^{-2}$) is taken into account. The dust column density however attenuates the FUV radiation. It is given by the optical depth $\tau$ using the conversion factor introduced in Sect. \ref{sec:uvdriven}. Figure \ref{fig:plot_uv_slides} shows the fractional abundance vs. the optical depth $\tau$. The main molecules CO, H$_2$O, CO$_2$, several PDR related species and the molecules showing the largest deviations are selected for Fig. \ref{fig:plot_uv_slides}. For most species only the results for a temperature of 550 K are shown, since FUV irradiation can considerably heat the gas through the photoelectric effect on dust grains. For reasons of clarity not all fractional abundances are given in this plot.\\

\begin{figure}[ht]
\plotone{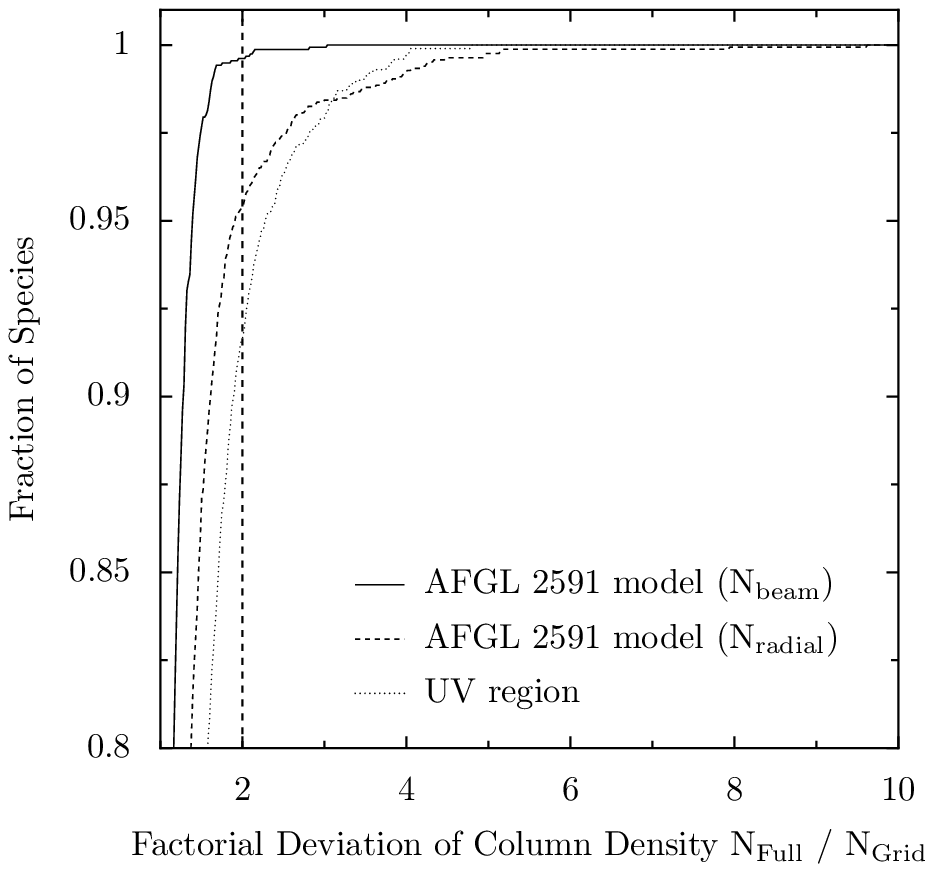}
\caption{Statistics on the accuracy of the interpolation method for chemical abundances: The fraction of species within a factorial deviation given in the X-axis is shown. Both benchmarks to the spherical AFGL 2591 model (Sect. \ref{sec:comp_afgl}) and the FUV model (Sect. \ref{sec:uvchem}) are given. For the AFGL 2591 model, the comparisons of radial and beam-averaged column densities are shown.\label{fig:compare_stat}}
\end{figure}

A value reflecting observable quantities is needed for the statistics comparing interpolated to fully calculated abundances (Fig. \ref{fig:compare_stat}). We use the column density of a molecule $i$ between $\tau=0.1$ and $\tau=15$ calculated by
\begin{equation}
N_i = 1.87 \times 10^{21} \ \int_{0.1}^{15} x_i(\tau) \, d\tau \ \ \ {\rm [cm}^{-2}{\rm ]} \ ,
\end{equation}
where $x_i(\tau)$ is the fractional abundance relative to the total gas density of the species $i$ depending on the optical depth $\tau$. All molecules in the chemical network are considered at five different values of $G_0$ and three temperatures. Only column densities larger than 10$^{11}$ cm$^{-2}$ are selected following the argument given in Sect. \ref{sec:comp_afgl}. A total of 2096 column densities are taken into account. The mean deviation for this benchmark is found to be $\left\langle Y_{\rm FUV} \right\rangle = 1.35$ and thus higher than the values in the previous section, but well within the goal of a factor of two.\\ 

%
%

Table \ref{tab:uvbads} in Appendix \ref{sec:dissagree} lists species, for which large factorial deviations in the column density between the grid and the fully calculated model were found. HSO$^+$ shows the largest deviation amounting to a factor of $Y=5.6$ for a high FUV flux of $G_0=10^6$ times the ISRF at a temperature of 100 K. The deviations can be explained by the sharp peak in fractional abundance at $\tau \approx 11$. It is a result of the formation through SO + H$_3^+$ $\rightarrow$ HSO$^+$ + H$_2$ or SO + HCO$^+$ $\rightarrow$ HSO$^+$ + CO. The peak in the abundance is also found in the fractional abundance of SO due to the competition of two FUV related reactions: At $\tau < 11$, SO is formed by the photodissociation of SO$_2$ and destroyed by the reaction with OH leading to SO$_2$. At larger $\tau$ however, SO dissociation becomes more important and its abundance thus decreases. This peak is not resolved by the current implementation of the grid as can easily be seen in a plot of the fractional abundance depending on $G_0$ and $\tau$ (Fig. \ref{fig:uv_grid_hso+_100k_1e6} in Appendix \ref{sec:dissagree}). While large deviations of HSO$^+$ occur at high $\tau$, the O$_2^+$ fractional abundance disagrees at low optical depth (Fig. \ref{fig:uv_grid_o2+_550k_1e6}).\\

What is the reason for these peaks in fractional abundance? Abundances in FUV irradiated regions are controlled by rate coefficients of the form $k \propto \exp\left( - \gamma \cdot \tau \right)$, where the product of the optical depth times a coefficient enters (see Eq. \ref{eq:uv_rate_fit}). Photodissociation rates thus drop exponentially with $\tau$. They become unimportant for the chemical abundance in a narrow interval of $\tau$. It should be noted however, that $\gamma$ is also affected by uncertainties at a level of at least $10\%$ due to the dust model used for fitting the rate and the range of $\tau$, over which the rates have been fitted (\citealt{vanDishoeck08b}).\\

\begin{figure}[ht]
\plotone{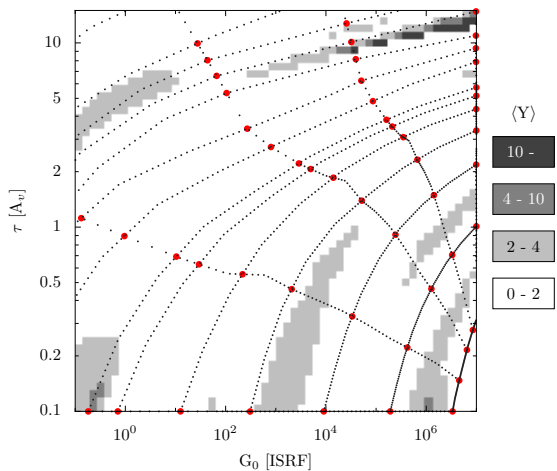}
\caption{Mean factorial deviation Y between the interpolated and fully calculated abundances of all species in the chemical network (grey scale). $G_0$ is the FUV field strength in units of the ISRF and $\tau$ the optical depth to the FUV source. The curvilinear coordinate system ($\alpha,\beta$) given in dotted line is used for the interpolation in the ($G_0,\tau$) - plane. The grid points are given by grey/red dots.\label{fig:plot_X_uv}}
\end{figure}

To reveal areas in the $(G_0,\tau)$ plane where deviations occur, the mean factorial deviation for all three temperatures and all species is shown in Fig. \ref{fig:plot_X_uv}. It shows that large areas agree well with $\langle Y \rangle < 2$, while other regions show a larger disagreement, however still mostly within a factor of 4. A narrow region in the $(G_0,\tau)$ plane is found to have mean deviations larger than a factor of 10. It coincidences with the values of $G_0$ and $\tau$ where a large deviation of HSO$^+$ has been found in a previous paragraph. As the statistics in Fig. \ref{fig:compare_stat} and the low mean deviation of the column density show, this local deviation is however sufficiently averaged out when observable quantities are derived from the grid.\\

%
%

%
%
\section{Utilizing the grid of chemical models} \label{sec:application}

The grid of chemical models is accessible through AMOUR (\textbf{A}bundance \textbf{M}odelling of young stellar \textbf{O}bjects \textbf{U}nder protostellar \textbf{R}adiation) available for public use at \anchor{http://www.astro.phys.ethz.ch/chemgrid.html}{http://www.astro.phys.ethz.ch/chemgrid.html}!. The page provides an online form to retrieve interpolated fractional abundances from the grid of chemical models. This form also calculates the X-ray ionization rate $\zeta_{{\rm H}_2}$ for a particular X-ray flux $F_{\rm X}$, X-ray attenuation $N({\rm H}_{\rm tot})$ and plasma temperature $T_{\rm X}$. For large modeling tasks or to include the interpolation method in other codes, the page also offers a FORTRAN code and the necessary molecular data tables in binary and ASCII format for download. Documentation is available at the website which describes the input/output parameters, the technical implementation, the format of the data tables, and shows example input/output data.\\

There exist a number of restrictions for the currently provided implementation which must be kept in mind when applying the method: The chemical composition in the molecular data table has been calculated for a specific chemical network (the UMIST 2006 database,\citealt{Woodall07}) using one set of initial conditions (Table \ref{tab:init_cond}). The current implementation does not include grain surface reactions except for H$_2$ formation on dust. The medium is assumed to be static and physical conditions (e.g., temperature or density) have been assumed to be constant over time. These are however only restrictions of the currently distributed molecular data tables. Further tables for other chemical networks, initial conditions, time-dependent parameters, etc. can be obtained from the authors on a collaborative basis.

%
%
\section{Conclusions and outlook} \label{sec:conclusion}

Starting from the chemical model of Doty et al. (2002, 2004) and St\"auber et al. (2004, 2005), we have introduced a new method to simulate the chemical evolution of YSO envelopes based on interpolation in a pre-calculated grid. The chemical model has been revised and the relevant physical parameters for the chemical composition of the gas are discussed. Benchmark tests have been carried out to verify the accuracy of the interpolation method. We conclude the following:

\begin{enumerate}
\item Accurate chemical modeling of a multidimensional envelope of YSOs is possible using a fast, grid-based interpolation method (Sect. \ref{sec:benchmark}). Comparison of observable quantities (beam averaged line fluxes) yields a mean factorial deviation between fully calculated and interpolated values of 1.35. This is more accurate than the uncertainties introduced by observations and chemical rate coefficients. Our method is more than five orders of magnitude faster than the full calculation (Sect. \ref{sec:multidimint}).
\item The X-ray models by \citet{Staeuber05} can be reproduced using an enhanced cosmic-ray ionization rate as a proxy for X-ray irradiation (Sect. \ref{sec:ionratxray}). In the relevant physical range for the application in YSO envelopes, the agreement in the fractional abundance is within $25\%$. The implementation of the this approach is described in detail in Appendix \ref{sec:app_xcross} and allows to include the effect of X-ray irradiation to chemical models in a simple way.
\item Ionization by X-rays and cosmic rays cannot be easily distinguished by molecular tracers. Spatial information on the abundance is thus needed to disentangle protostellar X-ray and cosmic-ray ionization (Sect. \ref{sec:ionratxray})
\item For the formation of CH$^+$ in low density gas (10$^4$ cm$^{-3}$) with high X-ray irradiation ($> 1$ erg $^{-1}$ cm$^{-2}$) recombination of doubly ionized carbon with \hh (C$^{++}$ + \hh $\rightarrow$ CH$^+$ + H$^+$) is important.
\item Increasing the initial abundance in the hot-core region ($T>100$ K) of the main sulphur carrier improves the agreement between models and observations in a high-mass star forming region (Sect. \ref{sec:ics_sulphur}).
\item Exploring the parameter space, we find regions with high gradients in molecular abundances. This is where small changes in the physical parameters yield large variations in abundance. As an example, interpolation of fractional abundances of an FUV irradiated region is difficult because photodissociation rates depend exponentially on the optical depth (Sect. \ref{sec:uvchem}). We have compensated for this by adopting a curvilinear coordinate system and a high number of grid points (Sect. \ref{sec:uvdriven}).
\end{enumerate}

The current grid method is limited by the assumption of the initial composition and fixed chemical rates. For different assumptions, the database needs to be recalculated. Using this fast chemical interpolation method, the radiative transfer calculation becomes the bottleneck in computing time to interpret data. Recently introduced escape probability methods, such as the exact method by \citet{Elitzur06} or an approximative multidimensional code by \citet{Poelman05} can speed up this step of the modeling.\\

This paper shows the possibility of interpolation for chemical modeling. In future publications, we will demonstrate major applications in multidimensional chemical modeling of YSOs and fast data-fitting to interpret observations. The gain in speed allows to carry out parameter studies on the influence of the geometry on the interpretation of observations. Furthermore, it will be possible to apply detailed chemical models to a large set of sources and draw conclusions on physical properties based on statistics. 

\acknowledgments
We thank Ewine van Dishoeck, Pascal St\"auber, Susanne Wampfler and an anonymus referee for useful discussions. Michiel Hogerheijde and Floris van der Tak are acknowledged for use of their RATRAN code. The submillimeter work at ETH is supported by the Swiss National Science Foundation grant 200020-113556 (SB and AOB). This work was partially supported under grants from The Research Corporation and the NASA grant NNX08AH28G (SDD).

\appendix

\section{A. Calculation of the ionization rate} \label{sec:app_xcross}

For the calculation of the X-ray ionization rate using the equation
\begin{equation} \label{eq:ion_zeta_h2_app}
\zeta_{{\rm H}_2} = \int_{E_{\rm min}}^{E_{\rm max}} J(E,r) \sigma_{\rm total}(E) \frac{E}{W(E) x({\rm H}_2)} dE \ \ ,
\end{equation}

we provide the compton and photoionization cross-sections as well as the mean energy $W(E)$ of an ion-pair in a tabulated form. The cross-sections are calculated using the X-ray model of \citet{Staeuber05} to which we refer for the source of the atomic and molecular constants and the exact implementation. We use the elemental composition given in Appendix \ref{sec:app_ics}. Since the photoionization cross-section of a molecule is approximated by adding the cross-sections of the contained atoms, the total photoionization cross-section does not depend on the abundance of the molecular species as long as the elemental composition does not change.\\
%
%
\begin{figure}[ht]
\begin{center}
\includegraphics[width=0.5\textwidth]{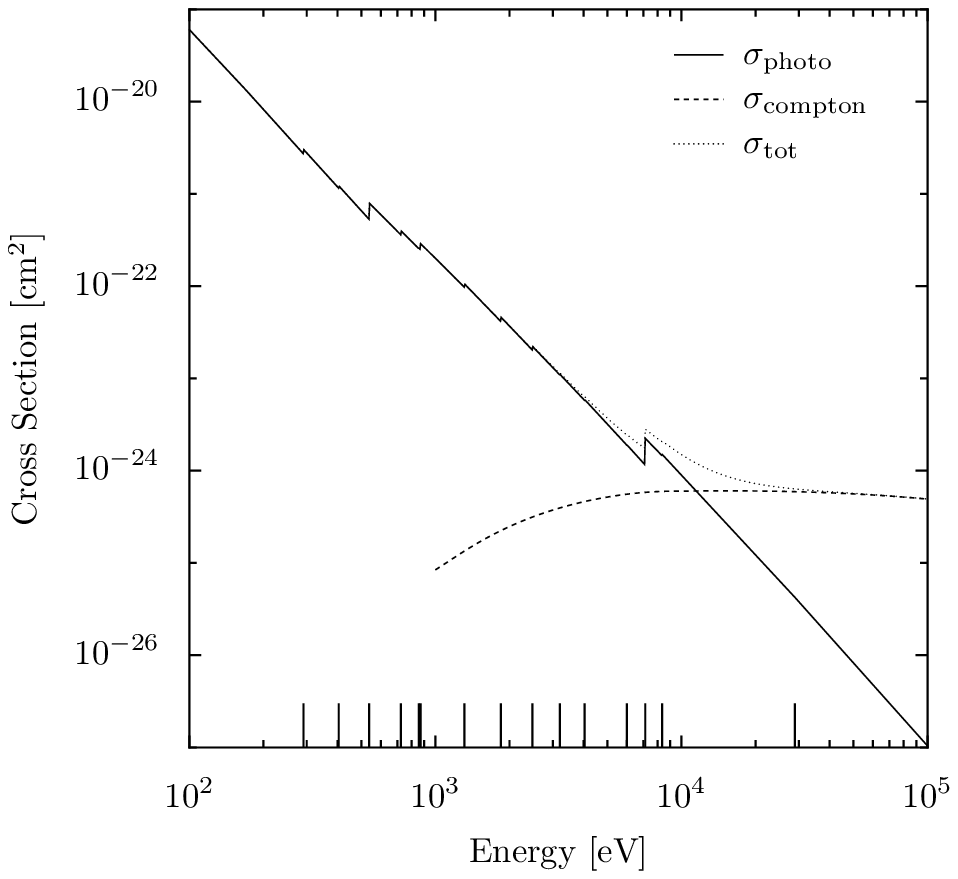}
\caption{Photoionization and compton cross-section versus photon energy. The ticks at the bottom of the figure indicate the supporting points for the interpolation of $\sigma$ given in Table \ref{tab:fit_cross}.\label{fig:plot_xcross}}
\end{center}
\end{figure}

The cross-sections are shown in Fig. \ref{fig:plot_xcross} for a photon energy between 100 and $10^5$~ eV. The photoionization cross-section, given by the solid line is approximated using a power-law, defined piecewise for an energy range between $E_{\rm min}$ and $E_{\rm max}$. Table \ref{tab:fit_cross} gives the cross-sections at the boundaries of each energy interval. The cross-section at an energy $E \in [E_{\rm min},E_{\rm max}]$ can then be calculated using
\begin{equation}
\sigma_{\rm photo}(E) \approx 10^{\log_{10}(\sigma_{\rm min}) \cdot (1-\alpha)+\log_{10}(\sigma_{\rm max}) \cdot \alpha} \ \ ,
\end{equation}
with $\alpha = (\log(E)-\log(E_{\rm min}))/(\log(E_{\rm max})-\log(E_{\rm min}))$. The deviation of the fit to the calculated cross-section is less than $5\%$ in the given energy range. Interpolation intervals are indicated by ticks at the bottom of the figure.\\

The cross-section for inelastic compton scattering is dominated by \hh and H. This process does not contribute to the total cross-section at low energy. We therefore use the fit from \citet{Staeuber05} to the XCOM database (NIST) for the energy above 1 keV. With $x=\log_{10}(E \ {\rm [eV]})$, the cross-section reads
\begin{eqnarray} \label{eq:sigmacompt}
\sigma_{\rm compton} &=& 2.869674 \cdot 10^{-23}-2.6364914 \cdot 10^{-23} \cdot x \\
                     &{}& +7.931175 \cdot 10^{-24} \cdot x^2-7.74014 \cdot 10^{-25} \cdot x^3 \nonumber \\	     
		     &{}& {\rm [cm}^2{\rm]} \ \ (E \leq 10 \, {\rm keV}) \nonumber \\		     
\sigma_{\rm compton} &=& -2.374894 \cdot 10^{-24}+1.423853 \cdot 10^{-24} \cdot x \nonumber \\
		     &{}& -1.70095 \cdot 10^{-25} \cdot x^2 \ \ \ \ {\rm [cm}^2{\rm ]} \ \ (E > 10 \, {\rm keV}) \nonumber
\end{eqnarray}

Finally, the mean energy per ion pair $W(E)$ is constant above 1 keV, $W(E > 1 {\rm [keV]}) = 20.95$~ eV, and can be approximated by 
\[
W(E)=23.65-(\log_{10}(E {\rm [eV]})-2) \cdot 2.7 \ \ {\rm  eV}
\]
between 100 and $10^3$~ eV, following \citet{Dalgarno99}.\\

\begin{table*}[tbh]
\begin{center}
\caption{Fitting parameters for the X-ray cross-section. a(b) means $a \times 10^b$.\label{tab:fit_cross}}
\begin{tabular}{c c c c}
\tableline\tableline
$E_{\rm min}$ [eV] & $E_{\rm max}$ [eV] & $\sigma_{\rm min}$ [cm$^2$]& $\sigma_{\rm max}$ [cm$^2$]\\
\tableline
$ 100   $  & $ 291   $    & $ 6.02(-20) $ & $ 2.71(-21) $ \\
$ 291   $  & $ 404.7 $    & $ 3.03(-21) $ & $ 1.15(-21) $ \\
$ 404.7 $  & $ 538   $    & $ 1.22(-21) $ & $ 5.29(-22) $ \\
$ 538   $  & $ 724   $    & $ 7.97(-22) $ & $ 3.59(-22) $ \\
$ 724   $  & $ 857   $    & $ 3.99(-22) $ & $ 2.54(-22) $ \\
$ 857   $  & $ 870   $    & $ 2.59(-22) $ & $ 2.48(-22) $ \\
$ 870   $  & $ 1311   $   & $ 2.89(-22) $ & $ 9.75(-23) $ \\
$ 1311   $ & $ 1846   $   & $ 1.06(-22) $ & $ 4.13(-23) $ \\
$ 1846   $ & $ 2477   $   & $ 4.61(-23) $ & $ 2.03(-23) $ \\
$ 2477   $ & $ 3203   $   & $ 2.23(-23) $ & $ 1.09(-23) $ \\
$ 3203   $ & $ 4043   $   & $ 1.11(-23) $ & $ 5.76(-24) $ \\
$ 4043   $ & $ 5996   $   & $ 5.89(-24) $ & $ 1.89(-24) $ \\
$ 5996   $ & $ 7124   $   & $ 1.91(-24) $ & $ 1.15(-24) $ \\
$ 7124   $ & $ 8348   $   & $ 2.24(-24) $ & $ 1.45(-24) $ \\
$ 8348   $ & $ 28900   $  & $ 1.50(-24) $ & $ 4.24(-26) $ \\
$ 28900   $& $ 100000   $ & $ 4.24(-26) $ & $ 1.04(-27) $ \\
\tableline
\end{tabular}
\end{center}
\end{table*}

\subsection*{The ionization rate of a thermal spectrum}

\begin{figure}[tbh]
\begin{center}
\includegraphics[width=0.5\textwidth]{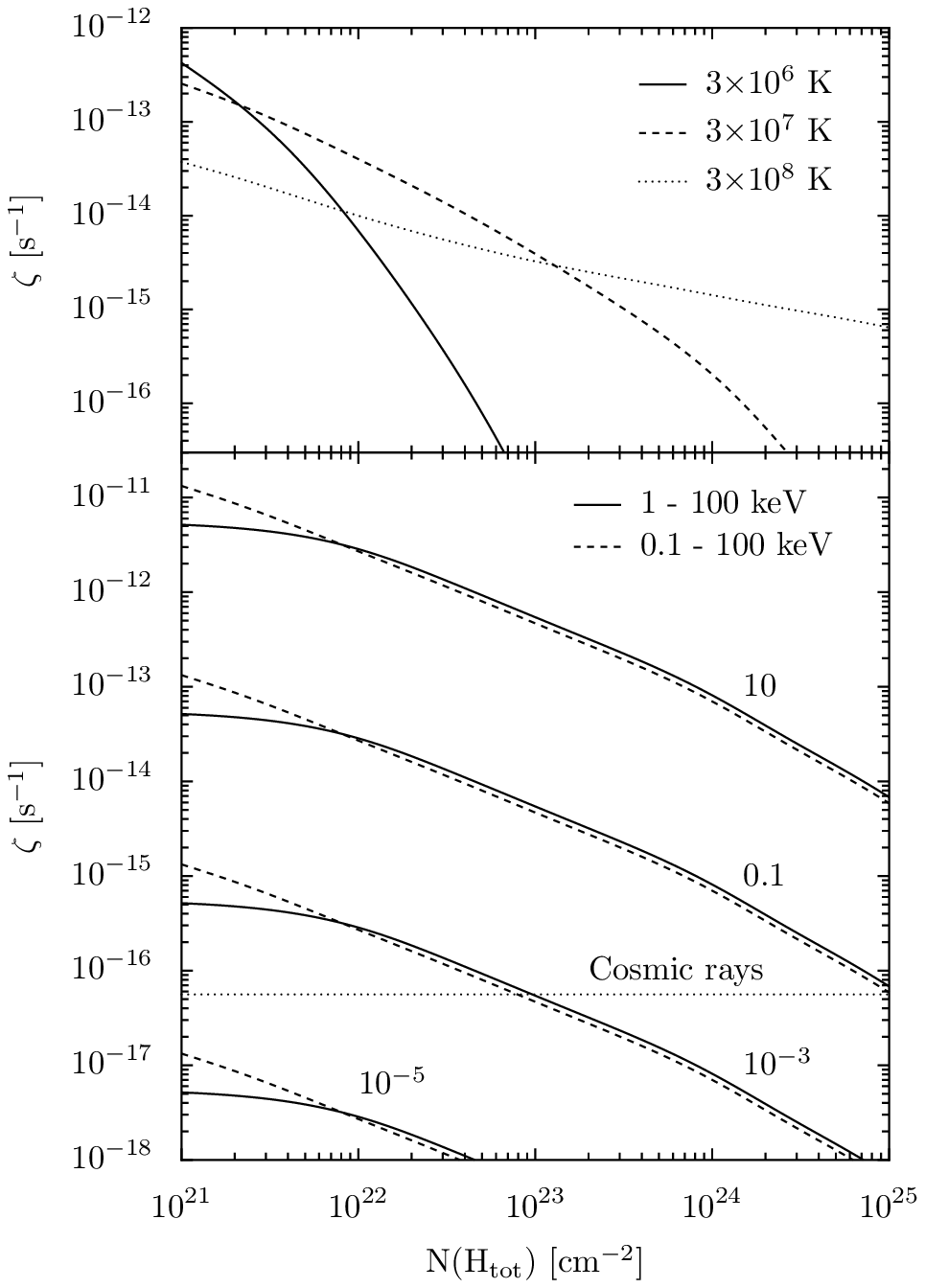}
\caption{Ionization rates assuming a thermal X-ray spectrum. The X-axis gives the attenuating column density. \textit{Top:} Ionization rate for different plasma temperatures for an X-ray flux of $10^{-2}$~ erg s$^{-1}$ cm$^{-2}$. \textit{Bottom:} Ionization rate for different X-ray fluxes ($10^{-5}, 10^{-3}, 0.1$ and 10 erg s$^{-1}$ cm$^{-2}$). Solid line: Eq. (\ref{eq:ion_zeta_h2_app}) is integrated from 1 - 100 keV. Dashed line: Eq. (\ref{eq:ion_zeta_h2_app}) is integrated from 0.1 - 100 keV.\label{fig:plot_xratesthermal}}
\end{center}
\end{figure}

As an important example, we calculate the ionization rates $\zeta$ [s$^{-1}$] for a thermal X-ray spectrum. The photo- and compton ionization cross-sections are implemented in the way described before and Eq. (\ref{eq:ion_zeta_h2_app}) is evaluated. Fig. \ref{fig:plot_xratesthermal} shows $\zeta$ depending on the attenuating column density. The ``standard'' cosmic-ray ionization rate of \tto{5.6}{-17} s$^{-1}$ is given by a dotted line. The top panel gives the ionization rate for 3 different plasma temperatures at a fixed flux of $10^{-2}$ erg s$^{-1}$ cm$^{-2}$. A hotter spectrum results in a larger number of photons at high energy which can penetrate further into the envelope. Most photons of a cold spectrum ($\approx10^6$~ K) however are quickly absorbed. For the bottom panel, the plasma temperature has been fixed to \tto{7}{7} K. Since the X-ray intensity depends linearly on the flux $F_{\rm X}$, the ionization rates scale in the same way. The difference in the ionization rate, when the integral in Eq. (\ref{eq:ion_zeta_h2_app}) is evaluated between 1-100 keV (solid line) and 0.1-100 keV (dashed line) is small at column densities larger than $10^{21}$ cm$^{-2}$, except for a small difference due to the normalization $\mathcal{N}$ (Eq. \ref{eq:x-flux}). Up to column densities of a few times $10^{21}$~ cm$^{-2}$, photons with an energy below 1 keV can however contribute significantly to the ionization rate.\\

In Table \ref{tab:fit_therm_xion}, the ionization rate for an X-ray flux of 1 erg s$^{-1}$ cm$^{-2}$ is given. The columns correspond to different plasma temperatures and the rows to different attenuating column densities. These ionization rates can be scaled linearly to an arbitrary value of the X-ray flux. For a point-like X-ray source, the flux at a distance $r$ from the source is obtained from the X-ray luminosity by $F_{\rm X}=L_{\rm X} / 4 \pi r^2$.

%
%

\begin{table*}[tbh]
\begin{center}
\caption{The ionization rate $\zeta_{{\rm H}_2}$ in s$^{-1}$ at $F_{\rm X} = 1$ erg s$^{-1}$ cm$^{-2}$. a(b) means $a \times 10^b$. The columns give values for different plasma temperatures [K] and the rows corresponds to column densities between $10^{20}$ and $10^{25}$ cm$^{-2}$.\label{tab:fit_therm_xion}}
\begin{tabular}{l c c c c c}
\tableline\tableline
$N({\rm H}_{\rm tot}$)& 3(6) K & 1(7) K & 3(7) K & 1(8) K & 3(8) K \\
\tableline
1(20) cm$^{-2}$ & 4.8(-11) & 2.4(-11) & 1.0(-11) & 3.4(-12) & 1.2(-12) \\
1(21) cm$^{-2}$ & 4.2(-12) & 4.5(-12) & 2.5(-12) & 9.8(-13) & 3.7(-13) \\
1(22) cm$^{-2}$ & 6.9(-14) & 3.7(-13) & 4.0(-13) & 2.2(-13) & 1.0(-13) \\
1(23) cm$^{-2}$ & 6.9(-17) & 9.6(-15) & 3.9(-14) & 4.5(-14) & 3.3(-14) \\
1(24) cm$^{-2}$ & 6.8(-22) & 4.5(-17) & 2.0(-15) & 9.4(-15) & 1.4(-14) \\
1(25) cm$^{-2}$ & 1.8(-30) & 8.6(-22) & 1.7(-17) & 1.4(-15) & 6.4(-15) \\
\tableline
\end{tabular}
\end{center}
\end{table*}

\clearpage

\section{B. Implementation of the UMIST 06 network} \label{sec:app_umist}

The UMIST 06 database for astrochemistry (\citealt{Woodall07}) lists an interval $[T_{\rm min}, T_{\rm max}]$ for the recommended temperature range for each reaction rate. For some reactions, different rates are given for distinct temperature ranges. The paper lists several rates, for which no extrapolation to low temperatures of $\approx 10$~ K should be done. We switch off those reactions for temperatures below $T_{\rm min}$.\\ 

All other reaction rates are extrapolated to lower temperatures with a few exceptions: Some reactions of the collider type (``CL'' in \citealt{Woodall07}) have a negative activation energy. To stabilize the chemical network, we keep their rates constant at temperatures below 500 K or 1000 K, or switch the reaction entirely off below 500 K (cf. Table \ref{tab:reac_temperature}).\\

Other reactions involve significant extrapolation, i.e., $T_{\rm min} > 300$~ K, and the rates do not decrease from 100 K to 10 K. Certain reactions are thus switched off outside their temperature range as given in the database. The reaction OH + CN $\rightarrow$ OCN + H is not switched off, since the OSU database\footnote{www.physics.ohio-state.edu/\~{}eric/research.html} (\citealt{Smith04}) lists this reaction with the same rate description. The reaction O + \hhco $\rightarrow$ CO + OH + H was introduced in UMIST 99 and has severe consequences for the abundance of formaldehyde: it is mainly destroyed by this reaction and the abundance drops by about two orders of magnitude at steady state conditions. The line flux, modeled with RATRAN \citep{Hogherheijde00} of a spherical model of AFGL 2591 is short by about two orders of magnitude compared to observations by \citet{vdTak99}. Thus we switch this reaction off below the recommended temperature range of 1750 K - 2575 K. Furthermore reactions with a rate more than an order of magnitude faster at 10 K than at 100 K are kept at constant below their minimum temperature, $k(T < T_{\rm min}) = k(T_{\rm min})$.

\begin{deluxetable}{llll}[h]
\tablecaption{Reactions not extrapolated to lower temperature in the implemented network (cf. text). The rates at 10 and 100 K are obtained using the UMIST 06 database. The number of the reaction in the UMIST database is given in the first column.\label{tab:reac_temperature}}
\tablehead{
\colhead{Nr.} & \colhead{Reaction} & \colhead{$k({\rm 10 K})$} & \colhead{$k({\rm 100 K})$} \\
\colhead{} & \colhead{} & \colhead{[cm$^{3}$ s$^{-1}$]} & \colhead{[cm$^{3}$ s$^{-1}$]}
}
\startdata
   158\tablenotemark{f} & CH + CH$_3$OH $\rightarrow$ CH$_3$ + H$_2$CO                   & 1.8(-7) & 2.1(-9) \\
   239\tablenotemark{d} & CH$_2$ + CH$_2$ $\rightarrow$ C$_2$H$_3$ + H                   & 3.3(-11) & 3.3(-11) \\
   242\tablenotemark{d} & CH$_2$ + O $\rightarrow$ CO + H$_2$                            & 8.0(-11) & 8.0(-11) \\
   243\tablenotemark{d} & CH$_2$ + O $\rightarrow$ HCO + H                               & 5.0(-11) & 5.0(-11) \\
   291\tablenotemark{d} & CH$_3$ + OH $\rightarrow$ H$_2$CO + H$_2$                      & 1.7(-12) & 1.7(-12) \\
   301\tablenotemark{d} & CH$_3$ + O$_2$ $\rightarrow$ HCO + H$_2$O                      & 1.7(-12) & 1.7(-12) \\
   338\tablenotemark{d} & O + H$_2$CO $\rightarrow$ CO + OH + H                          & 1.0(-10) & 1.0(-10) \\
   420\tablenotemark{d} & NH$_2$ + NO $\rightarrow$ N$_2$ + OH + H                       & 1.5(-12) & 1.5(-12) \\
   430\tablenotemark{e} & OH + CN $\rightarrow$ OCN + H                                  & 7.0(-11) & 7.0(-11) \\
   445\tablenotemark{d} & OH + HOCH $\rightarrow$ CO$_2$ + H$_2$ + H                     & 1.8(-11) & 1.8(-11) \\
   457\tablenotemark{f} & NH$_3$ + CN $\rightarrow$ HCN + NH$_2$                         & 2.0(-9) & 1.0(-10) \\
   491\tablenotemark{f} & CN + O$_2$ $\rightarrow$ OCN + O                               & 1.6(-9) & 3.2(-11) \\
   515\tablenotemark{d} & HCO + HCO $\rightarrow$ CO + CO + H$_2$                        & 3.6(-11) & 3.6(-11) \\
  4084\tablenotemark{f} & H + OH $\rightarrow$ H$_2$O + $\gamma$                         & 3.3(-14) & 6.6(-16) \\
  4552\tablenotemark{c} & CO + M $\rightarrow$ O + C + M                                 & $<$ 1.0(-30) & $<$ 1.0(-30) \\
  4554\tablenotemark{f} & H + CH$_3$ $\rightarrow$ CH$_4$ + M                            & 2.8(-26) & 4.5(-28) \\
  4555\tablenotemark{f} & H + O $\rightarrow$ OH + M                                     & 1.3(-30) & 1.3(-31) \\
  4558\tablenotemark{f} & H + OH $\rightarrow$ H$_2$O + M                                & 1.3(-28) & 1.4(-30) \\
  4563\tablenotemark{f} & H + O$_2$ $\rightarrow$ O$_2$H + M                             & 2.2(-28) & 1.2(-31) \\
  4568\tablenotemark{d} & H$_2$ + N $\rightarrow$ NH$_2$ + M                             & 1.0(-26) & 1.0(-26) \\
  4571\tablenotemark{b} & C + C $\rightarrow$ C$_2$ + M                                  & 1.9(197) & 1.7(-9) \\
  4573\tablenotemark{a} & C + O $\rightarrow$ CO + M                                     & 4.9(67) & 9.6(-19) \\
  4574\tablenotemark{a} & C$^+$ + O $\rightarrow$ CO$^+$ + M                             & 4.9(69) & 9.6(-17) \\
  4575\tablenotemark{a} & C + O$^+$ $\rightarrow$ CO$^+$ + M                             & 4.9(69) & 9.6(-17) \\
  4579\tablenotemark{a} & O + O $\rightarrow$ O$_2$ + M                                  & 3.2(-9) & 9.4(-32) \\
  4581\tablenotemark{f} & O + SO $\rightarrow$ SO$_2$ + M                                & 4.8(-28) & 6.9(-30) \\
  4582\tablenotemark{a} & OH + OH $\rightarrow$ H$_2$O$_2$ + M                           & 2.2(-2) & 1.9(-28) \\
  4583\tablenotemark{a} & O$_2$H + O$_2$H $\rightarrow$ H$_2$O$_2$ + O$_2$ + M           & 2.8(10) & 3.6(-29) \\
\enddata
\tablenotetext{a}{Kept constant below 500 K, $k(T < 500 {\rm K}) = k(500 {\rm K})$.}
\tablenotetext{b}{Kept constant below 1000 K, $k(T < 1000 {\rm K}) = k(1000 {\rm K})$.}
\tablenotetext{c}{Switched off below 500 K, $k(T < 500 {\rm K}) = 0$.}
\tablenotetext{d}{Switched off outside recommended temperature range, $k(T \not\in \left[T_{\rm min},T_{\rm max}\right]) = 0$.}
\tablenotetext{e}{Also listed in the OSU database, thus kept at recommended rate.}
\tablenotetext{f}{Kept constant below T$_{\rm min}$, $k(T < T_{\rm min}) = k(T_{\rm min})$.}
\end{deluxetable}

\clearpage

\section{C. Initial conditions} \label{sec:app_ics}

The total elemental abundances in Table \ref{tab:tot_elem} are used for the calculation of the photoionization cross-section, where also heavy elements locked into dust-grains are taken into account. The values are taken from \citet{Yan97}, except for helium, where the  values assumed by \citet{Staeuber05} have been adopted.\\

The initial conditions of species in the gas phase relative to the total hydrogen density ($n_{\rm tot}=2 n({\rm H}_2) + n({\rm H})$) are given in Table \ref{tab:init_cond}. If no other reference is given, we follow \citet{Doty02,Doty04} and \citet{Staeuber04,Staeuber05} for the abundances. All species in the network without specification are initially set to an absolute abundance of $10^{-8}$~ cm$^{-3}$ (effectively zero).\\

\begin{table*}[ht]
\begin{center}
\caption{Total elemental composition. a(b) means $a \times 10^b$.\label{tab:tot_elem}}
\begin{tabular}{l l| l l}
\tableline\tableline
Element & $n_{\rm tot}$(X)/$n_{\rm tot}$ & Element & $n_{\rm tot}$(X)/$n_{\rm tot}$\\
\tableline
H  & 1.0     & Fe & 3.2(-5)   \\
He & 8.5(-2) & Ne & 1.4(-4)   \\
C  & 3.5(-4) & Na & 2.1(-6)   \\
N  & 1.0(-4) & Mg & 4.0(-5)   \\
O  & 5.4(-4) & Al & 3.1(-6)   \\
S  & 2.0(-5) & Ar & 3.8(-6)   \\
P  & 1.0(-8) & Ca & 2.2(-6)   \\
Si & 3.5(-5) & Cr & 4.9(-7)   \\
Cl & 8.3(-8) & Ni & 1.8(-6)   \\
\tableline
\end{tabular}
\end{center}
\end{table*}

\begin{deluxetable}{lllc}[htb]
\tablecaption{Initial conditions in the gas-phase. a(b) means $a \times 10^b$.\label{tab:init_cond}}
\tablehead{
\colhead{Species} & \colhead{n$_{\rm Gas}$(X)/n$_{\rm tot}$} & & \colhead{Remark}}
\startdata
H$_2$      & 0.5     &               & \\
CO         & 1.8(-4) &               & \\
CO$_2$     & -       & $T < 100$ K   & \tablenotemark{a}\\
           & 1.5(-5) & $T \geq 100$ K& \\
H$_2$O     & -       & $T < 100$ K   & \tablenotemark{a}\\
           & 7.5(-5) & $T \geq 100$ K& \\
O          & 4(-5)   & $T < 100$ K   & \\
           & -       & $T \geq 100$ K& \tablenotemark{a}\\
H$_2$S     & -       & $T < 100$ K   & \tablenotemark{a}\\
           & 2(-5)   & $T \geq 100$ K& \tablenotemark{b}\\
S          & 9.1(-8) & $T < 100$ K   & \tablenotemark{b,c}\\
           & -       & $T \geq 100$ K& \tablenotemark{a}\\
N$_2$      & 3.5(-5) &               & \\
CH$_4$     & 5(-8)   &               & \\
C$_2$H$_4$ & 4(-8)   &               & \\
C$_2$H$_6$ & 5(-9)   &               & \\
H$_2$CO    & -       & $T < 60$ K    & \tablenotemark{a}\\
           & 6(-8)   & $T \geq 60$ K & \tablenotemark{d}\\
CH$_3$OH   & -       & $T < 60$ K    & \tablenotemark{a}\\
           & 5(-7)   & $T \geq 60$ K & \tablenotemark{d}\\
He         & 8.5(-2) &               & \\
He$^+$     & 2(-10)  &               & \tablenotemark{e}\\
H          & 5(-8)   &               & \tablenotemark{e}\\
H$^+$      & 3(-10)  &               & \tablenotemark{e}\\
H$_3^+$    & 2(-9)   &               & \tablenotemark{e}\\
HCO$^+$    & 3(-9)   &               & \tablenotemark{e}\\
H$_3$O$^+$ & 5(-10)  &               & \tablenotemark{e}\\
Grain$^-$  & 1.9(-8) &               & \tablenotemark{e,f}\\
e$^-$      & 7.5(-9) &               & \tablenotemark{e}\\
\enddata
\tablenotetext{a}{assumed to be frozen-out onto dust grains at cold temperatures or to be not abundant in hot regions.}
\tablenotetext{b}{see Sect. \ref{sec:ics_sulphur}}
\tablenotetext{c}{\citet{Aikawa08}}
\tablenotetext{d}{Evaporation temperature from \citet{Doty04}}
\tablenotetext{e}{\citet{Staeuber04,Staeuber05}}
\tablenotetext{f}{Negatively charged grains \citep{Maloney96}}
\end{deluxetable}

\clearpage

\section{D. Species with large disagreement} \label{sec:dissagree}

\begin{deluxetable}{lccccc}[htb]
\tablecaption{List of molecules with the largest factorial deviation of the grid interpolation from the fully calculated model found in benchmark tests in Sect. \ref{sec:benchmark}. The column densities of both approaches are given along with the factorial deviation Y.\label{tab:uvbads}}
\tablehead{
\colhead{Species} & \colhead{$N_{\rm full}$} & \colhead{$N_{\rm grid}$} & \colhead{$Y$} \\
\colhead{} & \colhead{[cm$^{-2}$]}    & \colhead{[cm$^{-2}$]}  \\
}
\startdata
\multicolumn{4}{l}{Sect. \ref{sec:comp_afgl}: AFGL 2591 model ($N_{\rm radial}$)} & $L_{\rm X}$ & Age \\
\multicolumn{4}{l}{} & [erg s$^{-1}$] & [yrs] \\
OH$^-$          &    1.9(10) &    1.8(11) &     9.6 & 10$^{32}$ & 3(3) \\
OH$^-$          &    1.1(11) &    8.9(11) &     7.9 & 10$^{31}$ & 3(3) \\
C$_6$H          &    3.2(12) &    6.2(11) &     5.2 & 0         & 5(4) \\
HCNH$^+$        &    6.4(13) &    1.2(13) &     5.1 & 10$^{30}$ & 3(5) \\
C$_7$H          &    2.1(12) &    4.2(11) &     5.0 & 0         & 5(4) \\
HCOOCH$_3$      &    5.7(11) &    1.1(11) &     4.9 & 0         & 5(4) \\
H$^-$           &    1.3(11) &    5.9(11) &     4.5 & 10$^{32}$ & 5(4) \\
HS$_2^+$        &    8.5(11) &    2.0(11) &     4.3 & 10$^{31}$ & 3(5) \\
H$^-$           &    1.4(11) &    6.0(11) &     4.3 & 10$^{32}$ & 3(5) \\
H$^-$           &    1.5(11) &    6.2(11) &     4.2 & 10$^{32}$ & 3(3) \\
CH$_2$CHCN      &    2.2(11) &    5.2(10) &     4.2 & 10$^{31}$ & 5(4) \\
HSO$_2^+$       &    4.9(12) &    1.2(12) &     4.0 & 0         & 3(5) \\
HS$_2^+$        &    8.5(11) &    2.1(11) &     4.1 & 10$^{31}$ & 5(4) \\
\tableline
\multicolumn{4}{l}{Sect. \ref{sec:comp_afgl}: AFGL 2591 model ($N_{\rm beam}$)} & $L_{\rm X}$ & Age \\
\multicolumn{4}{l}{} & [erg s$^{-1}$] & [yrs]\\
CH$_3$OH        &    8.3(12) &    2.7(12) &     3.0 & 0 &  5(4) \\
CH$_3$OCH$_3$   &    5.7(11) &    2.0(11) &     2.8 & 0 &  5(4) \\
H$_2$S          &    2.2(15) &    1.0(15) &     2.2 & 0 &  5(4) \\
HC$_7$N         &    3.3(11) &    1.6(11) &     2.1 & 0 &  5(4) \\
HSO$_2^+$       &    3.1(11) &    1.5(11) &     2.1 & 0 &  3(5) \\
S$_2$           &    1.4(12) &    2.8(12) &     2.0 & 0 &  3(5) \\
\tableline
\multicolumn{4}{l}{Sect. \ref{sec:uvchem}: FUV irradiated zone}  & $G_0$ & $T$\\
\multicolumn{4}{l}{}  & [ISRF]& [K]\\
HSO$^+$         &    3.0(12) &    5.3(11) &     5.6 &      $10^6$    &       101  \\
O$_2^+$         &    1.3(12) &    2.6(11) &     4.8 &      $10^2$    &       550  \\
CH$_3$OCH$_3$   &    2.5(12) &    6.1(11) &     4.1 &      $10^6$    &       101  \\
CH$_3$OH        &    5.3(12) &    1.3(12) &     4.0 &      $10^6$    &        80  \\
HC$_3$N         &    2.9(14) &    7.1(13) &     4.0 &      $10^6$    &       101  \\
C$_2$N          &    1.0(14) &    2.5(13) &     4.0 &      $10$      &       101  \\
\enddata
\end{deluxetable}

%
%
\begin{figure}[ht]
\begin{center}
\includegraphics[width=0.5\textwidth]{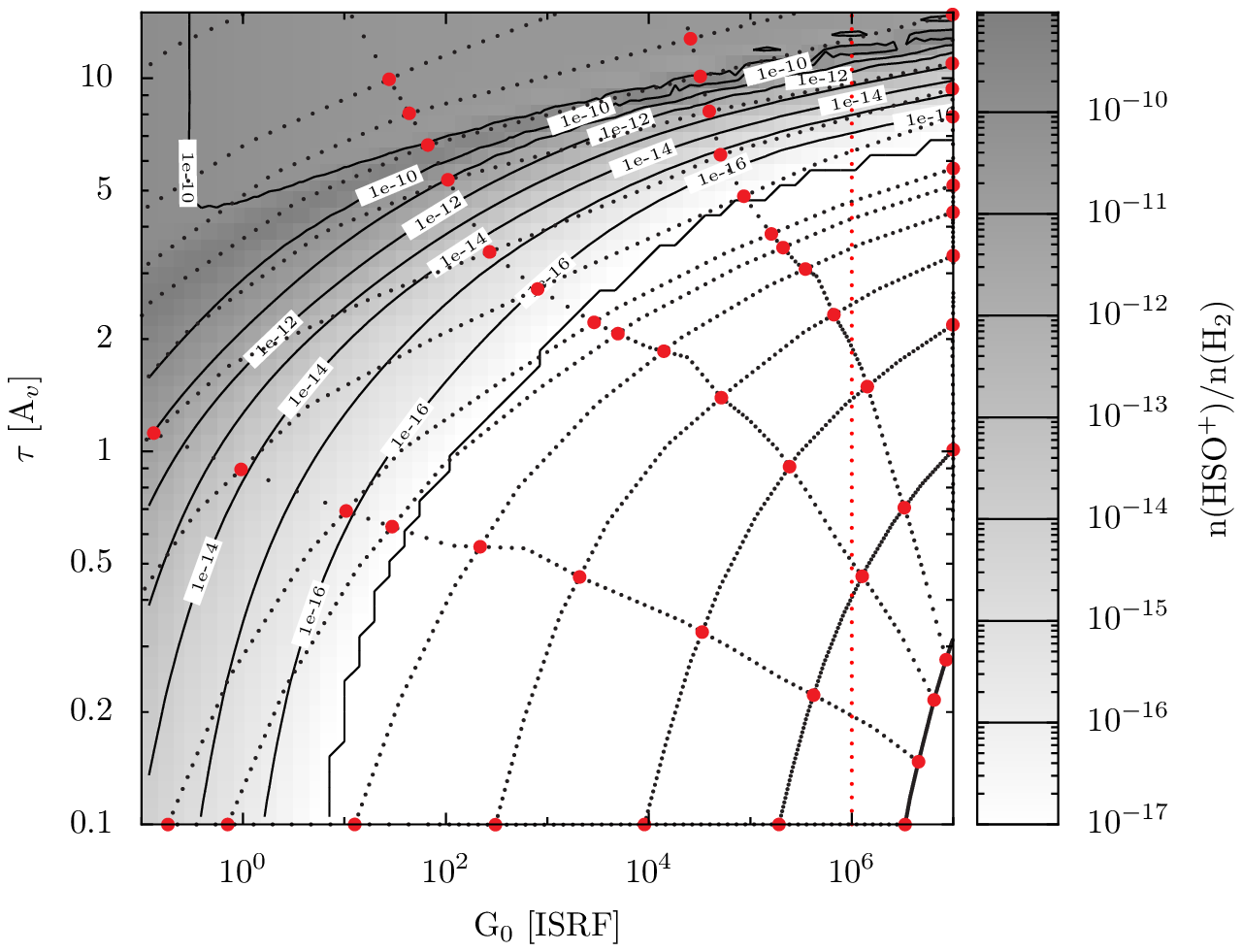}
\caption{Fractional abundance of HSO$^+$ (black solid line) for different FUV fluxes ($G_0$) and FUV attenuation  ($\tau$). The total density was chosen to be $10^6$ cm$^{-3}$, while the temperature is fixed at 100 K. The grid points of the curvilinear coordinate system are shown in (grey/red) dots and black dotted line. The value of $G_0$ ($10^6$) where large deviations were found is indicated by a vertical red/grey dotted line.\label{fig:uv_grid_hso+_100k_1e6}}
\end{center}
\end{figure}

%

%
\begin{figure}[ht]
\begin{center}
\includegraphics[width=0.5\textwidth]{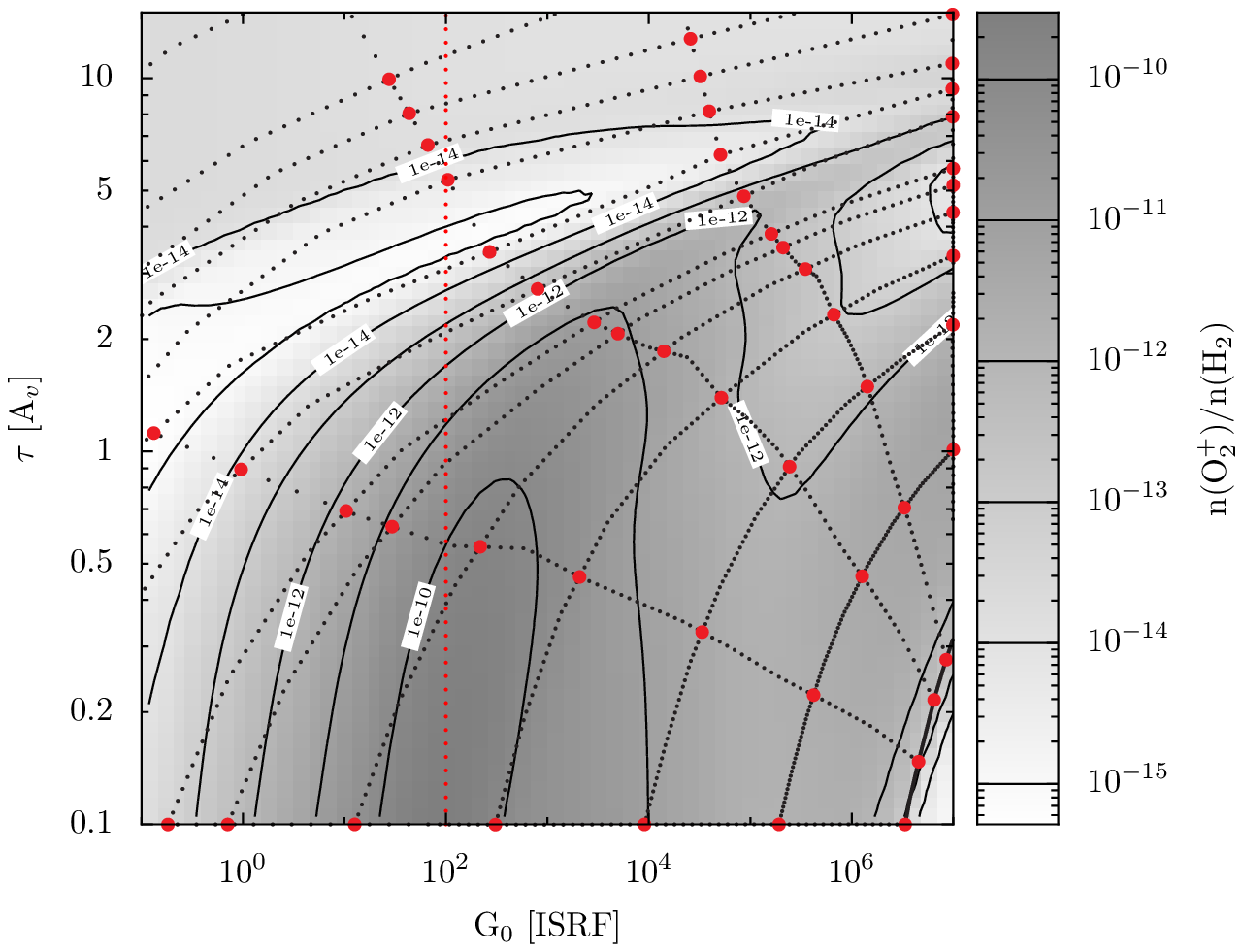}
\caption{Fractional abundance of O$_2^+$ (black solid line) for different FUV fluxes ($G_0$) and FUV attenuation  ($\tau$). The total density was chosen to be $10^6$ cm$^{-3}$, while the temperature is fixed at 550 K. The grid points of the curvilinear coordinate system are shown in (grey/red) dots and black dotted line. The value of $G_0$ ($10^2$) where large deviations were found is indicated by a vertical red/grey dotted line.\label{fig:uv_grid_o2+_550k_1e6}}
\end{center}
\end{figure}

\bibliographystyle{apj,apj-jour}

\clearpage

\end{document}